\documentclass[aps,pra,reprint,showpacs,floatfix,twocolumn]{revtex4-1}

\usepackage{amsmath,amsthm,amsfonts,amssymb,amscd}
\usepackage{t1enc}
\usepackage[utf8]{inputenc}
\usepackage[english]{babel}
\usepackage{babel}
\usepackage{graphics}
\usepackage{graphicx}
\usepackage[T1]{fontenc}
\usepackage{latexsym}
\usepackage{enumerate}
\usepackage[caption=false]{subfig}

\begin{document}

\title{Conditions for optical parametric oscillation with structured light pump} 
\author{G. B. Alves, R. F. Barros, D. S. Tasca, C. E. R. Souza, A. Z. Khoury}
\affiliation{Instituto de F\'isica, Universidade Federal Fluminense, CEP 24210-346, Niter\'oi-RJ, Brazil}
\begin{abstract}
We investigate the transverse mode structure of the down-converted beams generated by a type-II optical parametric oscillator (OPO) driven by a structured pump. Our analysis focus on the selection rules imposed by the spatial overlap between the transverse modes of the three fields involved in the non-linear interaction. These rules imply a hierarchy of oscillation thresholds that determine the possible transverse modes generated by the OPO, as remarkably confirmed with experimental results.
\end{abstract}
\maketitle

\section{Introduction}\label{sec-introduction}

For many years, the optical parametric oscillator (OPO) has attracted attention of the scientific community, partly due to the vast number of possibilities it brings to both fundamental and applied optics. In the literature, there are reports of OPOs being used for different purposes, such as in the detection of explosives \cite{ayrapetyan_laser_2018}, sensing of trace molecules \cite{vodopyanov_massively_2018}, ultrasonic testing of fiber reinforced plastics \cite{kusano_mid-ir_2018} and even for the generation of high power \textit{eye-safe} radiation \cite{kaskow_mw_2018}.  Besides, the OPO is a well-known source of non-classical states of light \cite{PhysRevLett.59.2555}, such as squeezed states, which have suppressed fluctuations in one of the field quadratures at the expense of increasing the noise in the other. In more recent works \cite{Mertz:91}, the noise suppression in the intensity difference between the generated beams has been measured down to 86\% below the shot-noise limit.

Another important feature of the OPO is that the signal and idler states it generates share EPR correlations \cite{PhysRevLett.68.3663,Keller:08}, which could be potentially useful for quantum information protocols with continuous variables \cite{braunstein2012quantum}, as attested by experimental realizations \cite{PhysRevLett.112.120505,PhysRevLett.107.030505}. In this regard, it has been shown that beams carrying orbital angular momentum (OAM) can be used in a number of protocols of quantum information, including teleportation  and quantum cryptography \cite{PhysRevA.77.032345,dambrosio_complete_2012}. Moreover, it has been shown that OAM entaglement is possible under spontaneous parametric down-conversion (SPDC) \cite{mair_entanglement_2001}, and also that entangled OAM states produced in an injected OPO hold equivalent properties as in a continuous-variable regime \cite{PhysRevLett.102.163602}.

The OAM of light, is an important property originated in the transverse structure of laser beams, which can be described in terms of the so-called paraxial modes, such as Laguerre-Gaussian (LG) or Hermite-Gaussian (HG), for example. These transverse modes are interesting for many different fields, from atomic physics \cite{Gahagan:96} to astrophysics \cite{Foo:05}. The spatial distribution of intensity noise have already been studied in connection with the transverse mode structure of semiconductor lasers \cite{Bramati:99,Hermier:99,892729}. Although pattern formation in the OPO dynamics has already been investigated long ago \cite{PhysRevA.49.2028,Marte:98,PhysRevLett.83.5278,PhysRevA.64.023803}, there are only a few studies on OPOs and cavity-free SPDC with transversely structured beams \cite{Schwob1998,PhysRevA.70.013812,Walborn:2012}. It has been shown, for example, that the three-mode coupling in the parametric down-conversion with LG modes imposes the OAM conservation between pump, signal and idler, and also that this OAM conservation may be broken in a type-II OPO due to anisotropies that cause the spectral separation of the different HG components of a given LG mode \cite{PhysRevA.70.013812,dosSantos:08}. However, the OPO dynamics under nontrivial mode structures is still a fruitful field of investigation \cite{Aadhi:17}.

In this paper, we present a detailed study of the OPO dynamics under different pump conditions and investigate the main features determining the spatial structure of the down-converted beams. We derive the selection rules for the transverse mode coupling and the corresponding oscillation threshold hierarchy based on the interaction strength between the transverse modes. 
Moreover, we address the role played by the cavity astigmatism introduced by the crystal's birrefringence, which breaks the frequency degeneracy and may prevent OAM conservation. General properties of the spatial modes generated by the OPO are then deduced from this analysis and remarkably confirmed through a variety of experimental data.

\section{Field propagation and cavity equations}

\begin{figure}[b]
\includegraphics[scale=0.95]{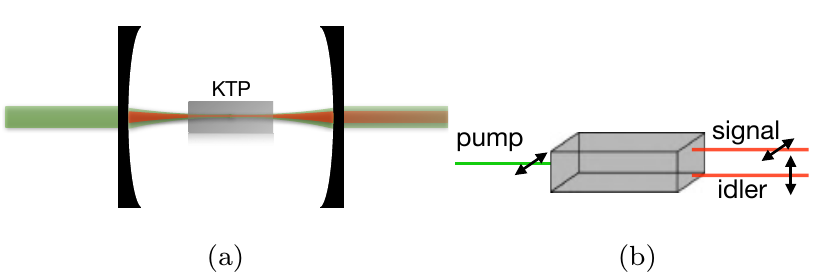}
\caption{Sketch of (a) the OPO cavity and (b) the non-linear medium ($\rm KTP - KTiOPO_4$) with the polarization of each field in a type-II configuration.}
\label{oi}
\end{figure}

In a non-magnetic nonlinear medium (crystal), the wave equation that describes the lossless propagation of light is
\begin{equation}
\nabla^2\mathbf{E}=\mu_0\epsilon\frac{\partial^2\mathbf{E}}{\partial t^2}+\mu_0\frac{\partial^2\mathbf{P}_{NL}}{\partial t^2} \,,
\label{eq-onda1}
\end{equation}
where $\mathbf{E}$ and $\mathbf{P}_{NL}$ are the electric field and the nonlinear polarization field, respectively. The non-linear process considered here is of second order where the non-linear polarization is proportional to the product of two fields, also known as three-wave mixing \cite{boyd2008nonlinear}. In this case, the wave equation has solution for three monochromatic waves with frequencies $\omega_0,\omega_1$ and $\omega_2$ such that $\omega_0=\omega_1+\omega_2$. Assuming a type-II configuration where each frequency $\omega_i$ has a fixed linear polarization \textit{i}, as shown in Fig. \ref{oi}, one has
\begin{equation}
E_i=\mbox{Re}\{E(\omega_i)\, e^{-i\omega_i t}\}
\label{eq-Ei}
\end{equation}
and
\begin{equation}
P_{NL,i}=\mbox{Re}\{P_{NL}(\omega_i)\, e^{-i\omega_i t}\}.
\label{eq-Pi}
\end{equation}
The frequency components of the non-linear polarization field can also be written in terms of the pump ($i=0$), signal ($i=1$) and idler ($i=2$) electric fields as
\begin{eqnarray}
P_{NL}(\omega_0)&=&d'E(\omega_1)E(\omega_2)  \label{p0}\\
P_{NL}(\omega_1)&=&d'E(\omega_0)E(\omega_2)^*\label{p1}\\
P_{NL}(\omega_2)&=&d'E(\omega_0)E(\omega_1)^*\label{p2}
\end{eqnarray}
where $d'$ is the second-order electric susceptibility. Substituting \eqref{eq-Ei}-\eqref{p2} in \eqref{eq-onda1}, we have
\begin{eqnarray}
\nabla^2 E(\omega_0)+k_0^2E(\omega_0)&=&-\mu_0\omega_0^2d'E(\omega_1)E(\omega_2) \label{eq-E0}\\
\nabla^2 E(\omega_1)+k_1^2E(\omega_1)&=&-\mu_0\omega_1^2d'E(\omega_0)E(\omega_2)^* \label{eq-E1}\\
\nabla^2 E(\omega_2)+k_2^2E(\omega_2)&=&-\mu_0\omega_2^2d'E(\omega_0)E(\omega_1)^* \label{eq-E2}
\end{eqnarray}
where $k_i=n_i \omega_i /c_0$ is the wave vector and $n_i$ is the refractive index.

Inside an optical resonator, this process is subjected to boundary conditions imposed by the cavity mirrors, resulting in a  discrete family of transverse modes. Thus, it is convenient to look for solutions for each propagating field as a superposition of paraxial modes $u_{pl}$, each of them with a \textit{z}-dependent  amplitude $\alpha_{pl}(z)$. In photon flux units, this superposition reads
\begin{equation}
E(\vec{r}\,;\omega_i)=\sqrt{\frac{2\hbar\omega_i}{n_i\epsilon_0 c_0}}\sum_{pl}u_{pl}^i(\vec{r})\,e^{ik_i z}\alpha_{pl}^i(z) \,,
\label{eq-decomp}
\end{equation}
where $\epsilon_0$ and $c_0$ are, respectively, the electric permittivity and the speed of light in free space, and modes $u_{pl}$ satisfy the so-called paraxial equation
\begin{equation}\label{eq-paraxial}
\nabla^2_\perp u_{pl}^{i} + 2ik_i\frac{\partial u_{pl}^{i}}{\partial z}=0\,.
\end{equation}

Substituting \eqref{eq-decomp} and \eqref{eq-paraxial} in \eqref{eq-E0}-\eqref{eq-E2}, one can easily show that the longitudinal evolution of the paraxial modes involved in a three-wave mixing process is described by the following set of equations \cite{Schwob1998}:
\begin{eqnarray}
\frac{d\alpha^0_{pl}}{dz}&=&i\chi\,e^{-i\Delta kz}\sum_{qm,\,rn}\left[\Lambda^{lmn}_{pqr}(z)\right]^*\alpha^1_{qm}\alpha^2_{rn} \label{eq-evol-1} \\
\frac{d\alpha^1_{qm}}{dz}&=&i\chi\,e^{i\Delta kz}\sum_{rn,\,pl}\Lambda^{lmn}_{pqr}(z)\alpha^0_{pl}\alpha^{2*}_{rn} \label{eq-evol-2}\\
\frac{d\alpha^2_{rn}}{dz}&=&i\chi\,e^{i\Delta kz}\sum_{qm,\,pl}\Lambda^{lmn}_{pqr}(z)\alpha^0_{pl}\alpha^{1*}_{qm} \label{eq-evol-3}
\end{eqnarray}
where
\begin{equation}\label{eq-recobrimento}
\Lambda^{lmn}_{pqr}(z)=\int\,d^2\vec{r}\,u^0_{pl}(\vec{r},z)u^{1*}_{qm}(\vec{r},z)u^{2*}_{rn}(\vec{r},z)\,,
\end{equation}
is a three-mode overlap integral which describes the coupling between the different transverse modes, $\Delta k= k_0-k_1-k_2$ is the so called phase-mismatch and $\chi$ is defined as
\begin{equation}
\chi=d'\sqrt{\frac{2\hbar\,\omega_0\omega_1\omega_2}{\epsilon_0^3c_0^3n_0n_1n_2}}\,.
\end{equation}
Equations  \eqref{eq-evol-1}-\eqref{eq-evol-3} describe the transverse mode coupling in the OPO.

By including the appropriate loss terms and computing the round trip variation of each mode amplitude, one derives the coarse-grained dynamical equations \cite{Schwob1998}
\begin{align}
&\frac{d\alpha^0_{pl}}{dt} = -\left(\gamma'_0-i\delta\varphi^0_{pl}\right)\alpha^0_{pl} + i\chi\sum_{qm,\,rn}(I^{lmn}_{pqr})^*\alpha^1_{qm}\alpha^2_{rn} \nonumber \\ &+t_0\alpha^{in}_{pl}\label{eq-OPO1} \\
&\frac{d\alpha^1_{qm}}{dt}=-\left(\gamma'_1-i\delta\varphi^1_{qm}\right)\alpha^1_{qm}+i\chi\sum_{pl,\,rn}I^{lmn}_{pqr}\,\alpha^0_{pl}\alpha^{2*}_{rn} \label{eq-OPO2} \\
&\frac{d\alpha^2_{rn}}{dt}=-\left(\gamma'_2-i\delta\varphi^2_{rn}\right)\alpha^2_{rn}+i\chi\sum_{pl,\,qm}I^{lmn}_{pqr}\,\alpha^0_{pl}\alpha^{1*}_{qm}\,, \label{eq-OPO3}
\end{align}
where $\gamma'_i$ represents the total cavity losses for the field $i$, $t_0$ is the transmissivity coefficient of the input mirror for the pump, the detuning $\delta\varphi^i_{mn}=\varphi^i_{mn}-2\pi q_i$ is the difference between the accumulated phase in a round trip and its value on the nearest cavity resonance and $t$ is the time measured in round trip units. The term $I^{lmn}_{pqr}$ represents an effective coupling constant and is defined as
\begin{equation}\label{eq-recobrimento-2}
I^{lmn}_{pqr}=\int_{-l/2}^{l/2} dz\,e^{i\Delta kz}\Lambda^{lmn}_{pqr}(z)\,,
\end{equation}
where $l$ is the crystal length.

\section{Selection rules for the transverse mode coupling}\label{sec-coupling}

\begin{figure*}[t!]
\centering
\includegraphics[width=0.6\textwidth]{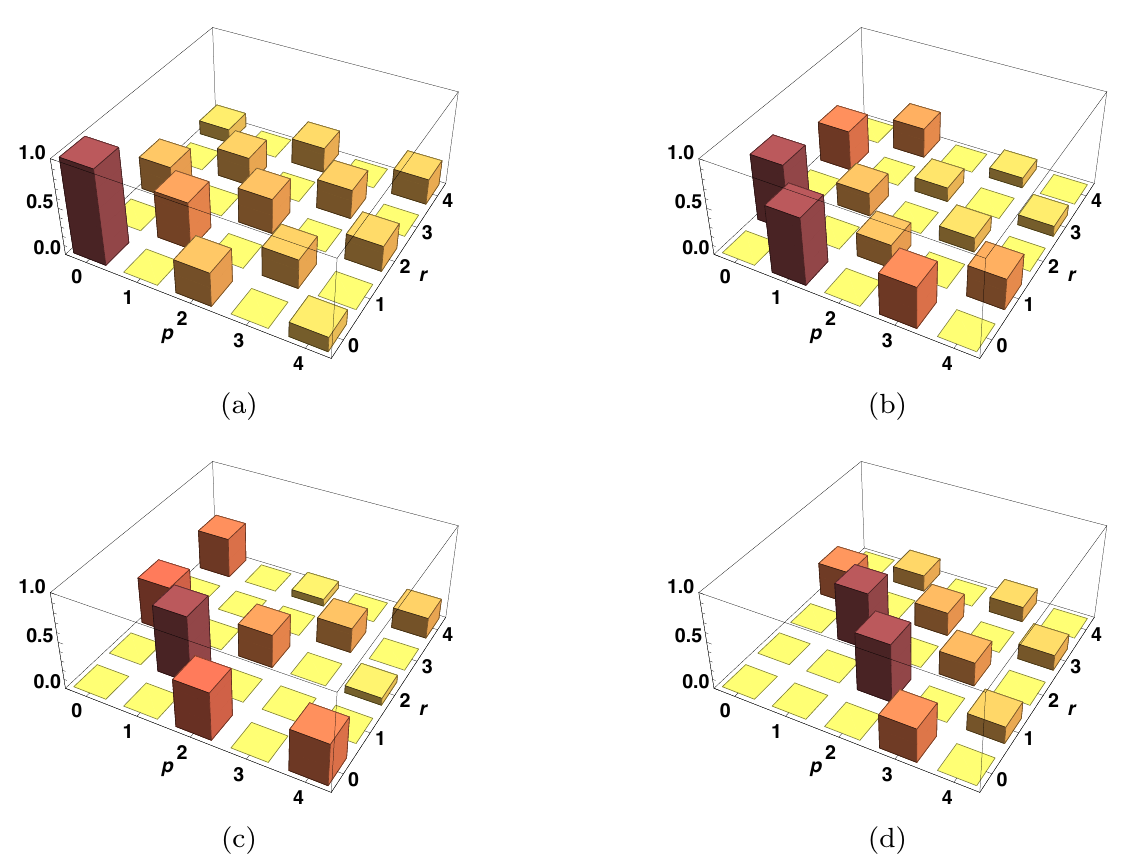}
\caption{Normalized magnitude of the coupling constant, $\Lambda^{mpr}_{000}/\Lambda^{000}_{000}=X^{mpr}$, in the Hermite-Gauss basis. The pump index $m$ varies for the different plots: $m=0$ (a), $m=1$ (b), $m=2$ (c) and $m=3$ (d).}
\label{fig-recobrimento-HG}
\end{figure*}

\begin{figure}
\centering
\includegraphics[width=0.3\textwidth]{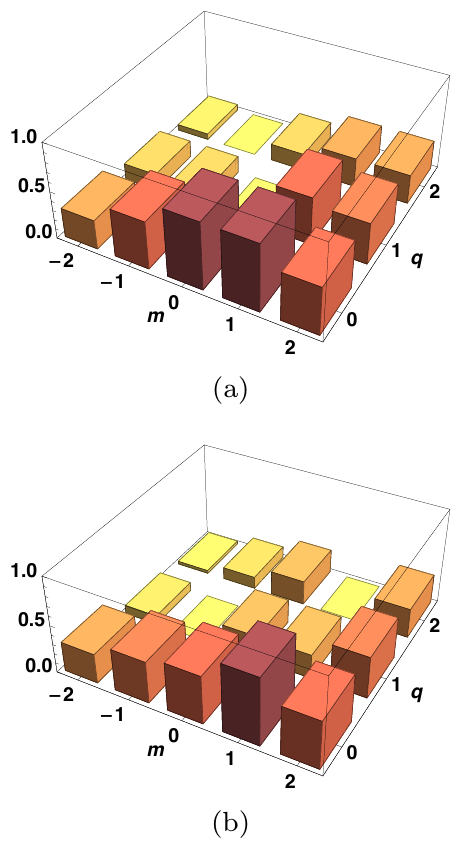}
\caption{Normalized magnitude of the coupling constant, $\Lambda^{lmn}_{pqr}/\Lambda^{000}_{000}$, in the Laguerre-Gauss basis. In (a), the pump mode has $\{l=1,p=0\}$ and we plot $\Lambda^{1m(1-m)}_{0q0}/\Lambda^{000}_{000}$. In (b), the pump mode has $\{l=2,p=0\}$ and we plot $\Lambda^{2m(2-m)}_{0q0}/\Lambda^{000}_{000}$.}
\label{fig-recobrimento-LG}
\end{figure}

In principle, the transverse mode dynamics involves an infinite set of coupled equations mediated by the coupling constants $I^{lmn}_{pqr}$. However, the spatial overlap and mode competition effects will restrict the non-vanishing transverse modes to a finite set. In this section, we derive the selection rules that result from spatial overlap and determine the non-zero coupling constants. We shall leave to section \ref{sec-hierarchy} the discussion about mode competition.

As mentioned before, the paraxial transverse modes $u_{pl}$ appearing in the expression \eqref{eq-recobrimento} can be conveniently chosen in either the LG or HG mode family. The relevant parameters of both families are presented in Appendix \ref{app-paraxial}. Firstly, we will work in the HG basis and then extend the result to the LG basis. Due to the geometry imposed by the cavity, all the three beams have (approximately) the same Rayleigh length $z_r$ \cite{Kogelnik:66}. Then, using expression (\ref{modo-HG}) for the HG modes in Eq.\eqref{eq-recobrimento}, after a straightforward calculation, we get

\begin{equation}\label{eq-recobrimento-4}
\begin{split}
&\Lambda^{mpr}_{nqs}(z)=C^{mpr}_{nqs}\frac{\gamma_1\gamma_2}{\pi}\sqrt{\frac{2}{\pi w_0^2(z)}}\frac{e^{i(S_1+S_2-S_0+1)\zeta(z)}}{\sqrt{1+z^2/z_r^2}}\times \\
&\int dx\, e^{-x^2}e^{\frac{\Delta k x^2}{2k_0}(1+\frac{iz}{z_r})}H_m(x)H_p(\gamma_1x)H_r(\gamma_2x)\times \\
&\int dy\, e^{-y^2}e^{\frac{\Delta k y^2}{2k_0}(1+\frac{iz}{z_r})}H_n(y)H_q(\gamma_1y)H_s(\gamma_2y)
\end{split}
\end{equation}
where $C^{mpr}_{nqs}=\sqrt{\tfrac{2^{-(m+n+p+q+r+s)}}{m!n!p!q!r!s!}}$, $\zeta(z)=\tan^ {-1}(z/z_r)$, $\gamma_i=w_0/w_i$ is the ratio between the pump ($w_0$) and down-converted ($w_i$) beam waists, $S_i$ is the paraxial mode order of field $i$ ($S_0=m+n$, $S_1=p+q$, $S_2=r+s$. See Appendix \ref{app-paraxial}) and $x$ ($y$) is a normalized transverse coordinate defined by the change of variables $\tfrac{\sqrt{2}x}{w_0(z)}\mapsto x$ ($\tfrac{\sqrt{2}y}{w_0(z)}\mapsto y$). We assume $\Delta k\ll k_0$ and $l\lesssim z_r$, which is frequently the case for the experimental conditions. Besides, the Gaussian terms restrict the relevant contributions of the integrands in \eqref{eq-recobrimento-4} to the region $|x|\lesssim 1$ and $|y|\lesssim 1$. Under these assumptions, the coupling constant can be factorized as the product of a purely longitudinal factor and a transverse overlap integral
\begin{equation}
\Lambda^{mpr}_{nqs}(z)\simeq \Lambda^{mpr}_{nqs}(0)\frac{e^{i(S_1+S_2-S_0+1)\zeta(z)}}{\sqrt{1+z^2/z_r^2}}\,,
\end{equation}
where
\begin{equation}\label{eq-recobrimento-5}
\begin{split}
&\Lambda^{mpr}_{nqs}(0)=C^{mpr}_{nqs}\frac{\gamma_1\gamma_2}{\pi}\sqrt{\frac{2}{\pi w_0^2}} \\
&\int dx\, e^{-x^2}H_m(x)H_p(\gamma_1x)H_r(\gamma_2x)\times \\
&\int dy\, e^{-y^2}H_n(y)H_q(\gamma_1y)H_s(\gamma_2y)\,.
\end{split}
\end{equation}

Note that the parity of the three Hermite polynomials product is equal to $m+p+r$ in the integral over the $x$ coordinate. Since $e^{-x^2}$ is an even function, then one must have $m+p+r = 0\;(\rm mod\:2)$, otherwise the integral would be zero. Furthermore, we can expand the product of the last two Hermite polynomials as a combination of other polynomials of the same parity:
\begin{equation}\label{eq-hphr}
H_p\left(\gamma_1 x\right)H_r\left(\gamma_2 x\right)=\sum_{i=0}^{\lfloor (p+r)/2\rfloor}\beta_{p,r,i}\,H_{p+r-2i}(x)\,.
\end{equation}
Using the orthogonality property of the polynomials, and substituting Eq.\eqref{eq-hphr} in \eqref{eq-recobrimento-5} results in
\begin{equation}
\begin{split}
&\sum_{i=0}^{\lfloor (p+r)/2\rfloor}\beta_{p,r,i}\int^{+\infty}_{-\infty}dx\, H_m\left(x\right)H_{p+r-2i}(x)\, e^{-x^2}=\\
&\sum_{i=0}^{\lfloor (p+r)/2\rfloor}\beta_{p,r,i}\,\delta_{m,p+r-2i} \,,
\end{split}
\end{equation}
which means that $m\leq p+r$. Analogously, from the $y$ integration, we obtain $n+q+s = 0\;(\rm mod\,2)$ and $n\leq q+s$. These relations imply that
\begin{eqnarray}
&S_0+S_1+S_2\equiv 0 \,\,(\rm mod\: 2)& \label{eq-conditions-1} \\
&S_0\leq S_1+S_2 & \label{eq-conditions-2}
\end{eqnarray}
which are basis independent conditions. These are general constraints regarding the transverse mode coupling in three-wave mixing.

Exact expressions for the transverse overlap Eq.(\ref{eq-recobrimento-5}) are provided in Appendix \ref{app-coupling}, from which one can derive the following expression for the effective coupling constant
\begin{equation}\label{eq-recobrimento-tot}
I^{mpr}_{nqs}=z_r\left(\int^{l/2z_r}_{-l/2z_r} du \, \frac{e^{i(\Delta k z_r)u}}{1-iu}\left[\frac{1+iu}{\sqrt{1+u^2}}\right]^{\Delta S}\right)\Lambda^{mpr}_{nqs}(0)\,.
\end{equation}
The term between parentheses can be viewed as an overall conversion efficiency, since it does not make explicit reference to the particular modes being coupled, except for the mode order difference $\Delta S$.

We can have an idea of the relative magnitude of the different coupling coefficients by looking at Figs. \ref{fig-recobrimento-HG} and \ref{fig-recobrimento-LG}. In those plots, the input pump indexes ($\{m,n\}$ for HG, or $\{l,p\}$ for LG modes) are kept fixed while the signal and idler indexes are unconstrained, as in the case of spontaneously down-converted beams. It can be observed that the order is conserved ($S_1+S_2=S_0$) for the highest coupling magnitudes.

\begin{figure}[h!]
\includegraphics[width=0.45\textwidth]{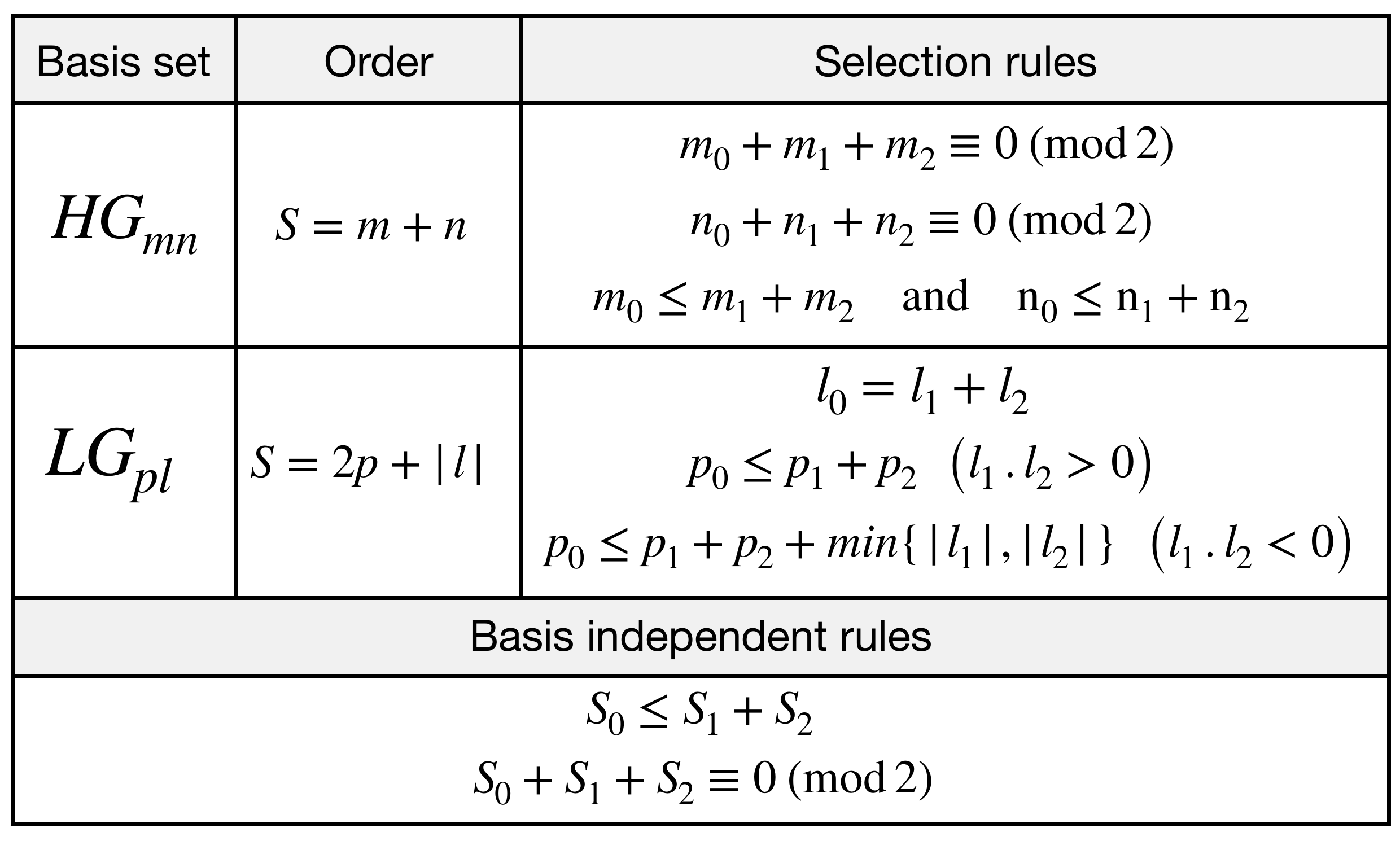}
\caption{A summary of the selection rules.}
\label{table-rules}
\end{figure}

In the Hermite-Gauss basis, one can see from Eq.\eqref{eq-recobrimento-5} that $\Lambda^{mpr}_{nqs}(0)$ can be broken into a product of two terms: $\Lambda^{mpr}_{nqs}(0)=X^{mpr}Y_{nqs}$, one related to the indexes in the $x$ coordinate, and the other to the $y$ coordinate. Since the functions $X^{mpr}$ and $Y_{nqs}$ are the same, one can analyze the maximization conditions for $X^{mpr}$ and then extend the result to $Y_{nqs}$. As seen from Fig.\ref{fig-recobrimento-HG}, when $p+r=m$, with $p=\lfloor m/2\rfloor$ and $r=\lceil m/2\rceil$ (or vice-versa), maximum mode coupling is attained. An analogous conclusion can be drawn for the $Y_{nqs}$ function and, recalling that $S_1=p+q$ and $S_2=r+s$, it confirms the \textit{order conservation} statement claimed before. This result is then independent of the basis used to describe the coupling and should be valid for the Laguerre-Gauss basis as well. However, in this case the topological charge conservation ($l_0=l_1+l_2$) \cite{Schwob1998} should also be taken into account, what makes Eq.\eqref{eq-conditions-1} automatically satisfied. Furthermore Eq.\eqref{eq-conditions-2} will introduce restrictions on the radial indexes of the modes, depending on the relative signs of the topological charges being added. This kind of cross-talk between radial and angular degrees of freedom has already been investigated in nonlinear OAM mixing \cite{PhysRevA.96.053856,Buono:18}. All these results are summarized in Fig.\ref{table-rules} for the different paraxial bases. The relative magnitudes of the different coupling constants will be essential for the mode selection in the OPO dynamics, as we shall see in section \ref{sec-results}.

\section{Experimental setup}\label{sec-exp-setup}

\begin{figure}[h]
\includegraphics[width=0.45\textwidth]{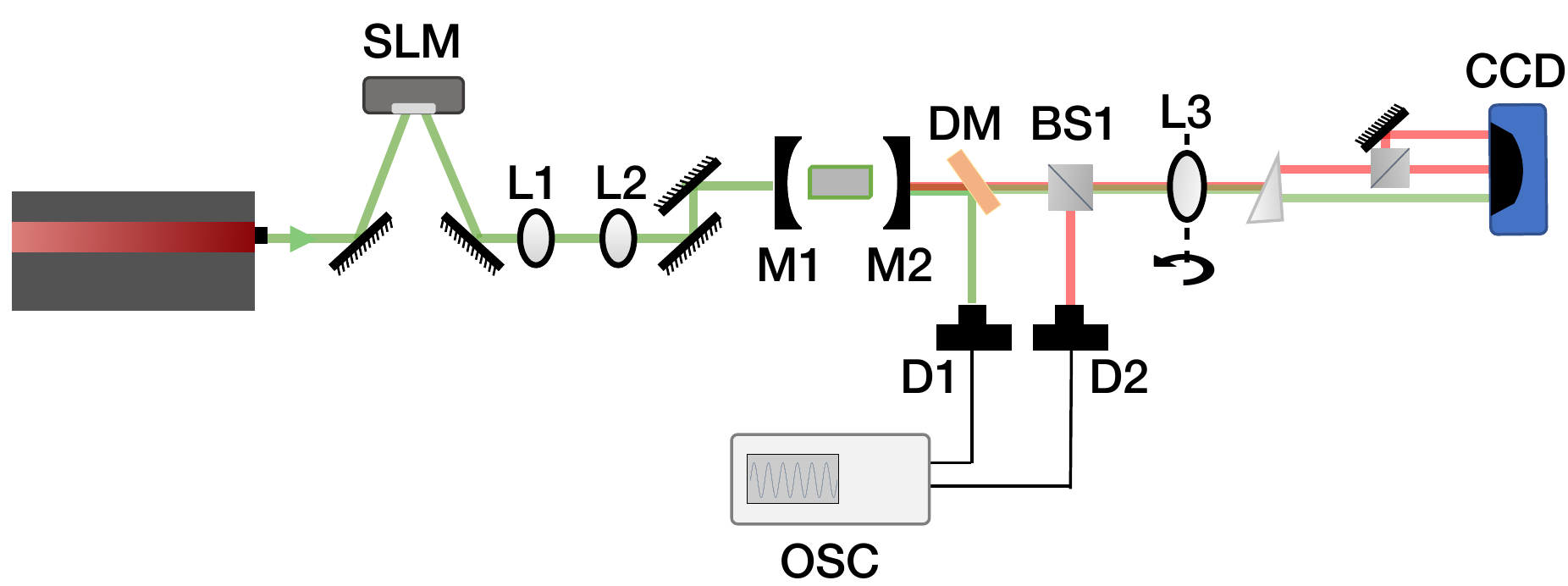}
\caption{A schematic view of the experiment. Lenses L1 and L2 are used to adjust the waist of the structured beam after the SLM, in order to pump the OPO cavity with a properly matched mode.}
\label{fig-experiment}
\end{figure}

The experimental setup is depicted in Fig.\ref{fig-experiment}. A 532nm $\rm TEM_{00}$ beam, originated from the second harmonic of a Nd:YAG laser (InnoLight GmbH laser, Diabolo product line), is collimated and sent towards a Spatial Light Modulator (SLM - Hamamatsu, model X10468-01). The SLM is electronically programmed to transform the beam into different paraxial higher-order modes, implemented by an amplitude and phase modulation technique, as described by method A in Clark \textit{et al.} \cite{Clark:16}. 

The OPO consists of two concave mirrors (M1 and M2) with equal radii of $R=25$mm and a 5mm long KTP ($\rm KTiOPO_4$) crystal, cut for 532-1064nm type-II phase matching at room temperature. The cavity is kept nearly confocal, while its length is controlled by a piezoelectric ceramic coupled to M2. The mirror M1 has a reflectance of 96\% at 532nm and is highly reflective (HR, R=99.8\%) at 1064nm, while M2 is HR at both wavelengths. The crystal can also be positioned and oriented with translation and rotation stages with micrometric precision.

The outcoming (resonant) beams pass through a dichroic mirror (DM) which directs most of the pump intensity to the detector D1 (Thorlabs, DET100A), used for monitoring the 532nm resonances in the oscilloscope (OSC) as the piezoelectric actuator scans the cavity. The transmitted beams are split in a 50/50 non-polarizing cube (BS1), which allows the analysis of the down-converted beams detected at D2. At last, the (residual) pump beam is spectrally separated by a prism, while signal and idler are separated by a PBS, allowing the simultaneous imaging of the three fields at the CCD (charge coupled device) camera. In order to identify the topological charges of the Laguerre-Gaussian (LG) modes, all beams pass through a lens L3 (f=200mm), which can be tilted to achieve LG to HG conversion \cite{VAITY20131154} right at the CCD sensor.

Because of the damage threshold of the SLM, the laser's output power is limited to 100mW, which means a maximum pump power of approximately 30mW (due to the SLM efficiency and other optical losses). Despite the relatively low pump intensity, the low reflection and absorption losses in the mirrors and crystal produce a high finesse cavity and an OPO with a relatively low oscillation threshold (down to approx 3mW for the $\rm TEM_{00}$ pump). At maximum power, we were able to achieve oscillation for pump transverse modes up to 3rd order. Hence, this experimental arrangement allowed us to characterize the transverse resonances for the signal and idler beams with respect to the pump beam in the case of a triply resonant optical parametric oscillator.

\section{Results and analysis}\label{sec-results}

In the experiment described in section \ref{sec-exp-setup}, we pumped the OPO with a variety of LG and HG modes and registered the various transverse modes populated by the down converted photons. We applied slight changes in the crystal orientation and cavity length in order to geometrically tune various transverse mode resonances for signal and idler. It is important to keep in mind that, in such optical setups with a nonlinear crystal placed inside an optical cavity, the resonant transverse modes possess an astigmatic Gouy phase, caused by the crystal birrefringence \cite{PhysRevA.70.013812}, resulting in a splitting of the resonant positions even for modes of the same order. For our case of a type-II conversion and a high finesse cavity for the down converted beams ($\gamma'\approx5$ mrad), it means that the ordinary polarization cannot encompass simultaneous oscillation for different Hermite-Gauss components ($|\Phi^2_{m+1,n-1}-\Phi^2_{m,n}|\approx 21$ mrad), and, as a corollary, the idler beam will oscillate in a pure Hermite-Gauss mode. Therefore, the Hermite-Gaussian basis is the most natural one for the type of anisotropy experienced by the interacting fields inside the OPO cavity.

\begin{figure}
\centering
\includegraphics[width=0.4\textwidth]{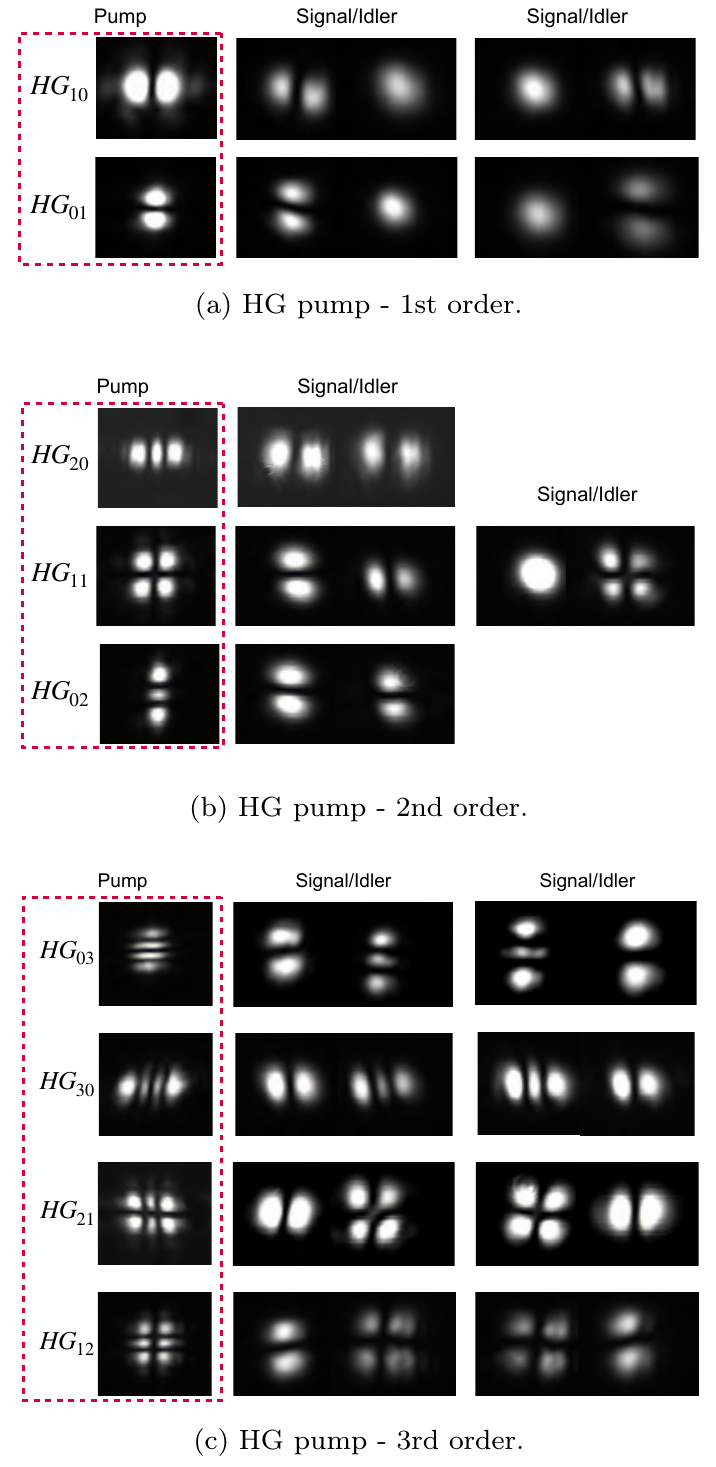}
\caption{Experimental images for the pump, signal and idler beams, corresponding to different Hermite-Gaussian orders of the pump mode.}
\label{fig-pump-HG}
\end{figure}

One key point for the transverse mode selection in the OPO operation is to note that the oscillation threshold decreases with increasing coupling strength: $|\alpha_{pl}^{in}|^2_{th}\sim |\Lambda^{lmn}_{pqr}|^{-2}$  \cite{debuisschert_type-ii_1993}. Since $|\Lambda^{000}_{000}|$ is the greatest coupling constant, the triply resonant Gaussian mode operation for pump, signal and idler has the lowest oscillation threshold. However, a more careful analysis is required when higher order pump modes and anisotropy are present. As a first illustration of the transverse mode selection rules, we present several experimental results obtained with different Hermite-Gauss pump modes. These results are shown in Fig.\ref{fig-pump-HG}. All results are compatible with the optimal coupling predicted in Fig.\ref{fig-recobrimento-HG}. Nevertheless, a more involved dynamics takes place when multiple Hermite-Gaussian modes are simultaneously present in the pump beam, as is the case when optical vortices and OAM transfer are considered.  

\begin{figure}[h]
\centering
\includegraphics[width=0.3\textwidth]{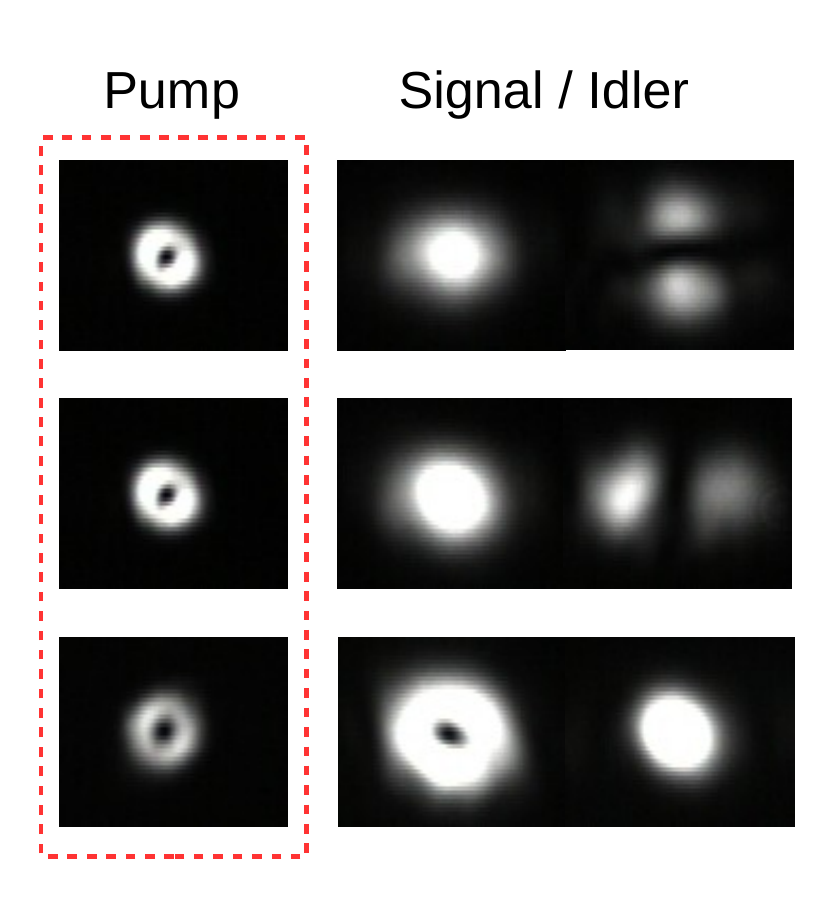}
\caption{Experimental images for the pump, signal and idler beams. The pump field is kept in the $u^{LG}_{01}$ mode.}
\label{fig-pump-01}
\end{figure}

For example, with a first order LG ($\{l=1,p=0\}$) pump in the anisotropic cavity, different operation conditions can be observed by tuning the cavity parameters. This can be analyzed through the pump decomposition in the HG modes
\begin{equation}
u^{LG}_{01}=\frac{1}{\sqrt{2}}\left(u^{HG}_{10}-iu^{HG}_{01}\right)\,.
\end{equation}
Hence, each Hermite-Gaussian component will couple with the transverse modes with the optimal overlap with it. As described above for the Hermite-Gauss basis, they are given by $\Lambda^{110}_{000}$ (or $\Lambda^{101}_{000}$) and $\Lambda^{000}_{110}$ (or $\Lambda^{000}_{101}$), giving rise to the following processes
\begin{eqnarray}
u^{HG}_{10}&\rightarrow &u^{HG}_{10}+u^{HG}_{00}\,, \label{eq-p1} \\
u^{HG}_{01}&\rightarrow &u^{HG}_{01}+u^{HG}_{00}\,. \label{eq-p2}
\end{eqnarray}
Since the idler has to oscillate in a pure HG mode, there are two possibilities:
\begin{enumerate}
\item idler in either $u^{HG}_{10}$ or $u^{HG}_{01}$: in this case, only one process is activated and the signal is necessarily in the $u^{HG}_{00}$ mode. This kind of operation prevents any OAM transfer to the signal. These cases are represented in the first two rows of Fig.\ref{fig-pump-01}.
\item idler in $u^{HG}_{00}$: in this case, the two processes \eqref{eq-p1} and \eqref{eq-p2} can be activated, allowing for the signal mode to be in a superposition of $u^{HG}_{10}$ and $u^{HG}_{01}$, with partial or total OAM transfer. This situation is shown in the last row of Fig.\ref{fig-pump-01}.
\end{enumerate}

\begin{figure}
\centering
\includegraphics[width=0.3\textwidth]{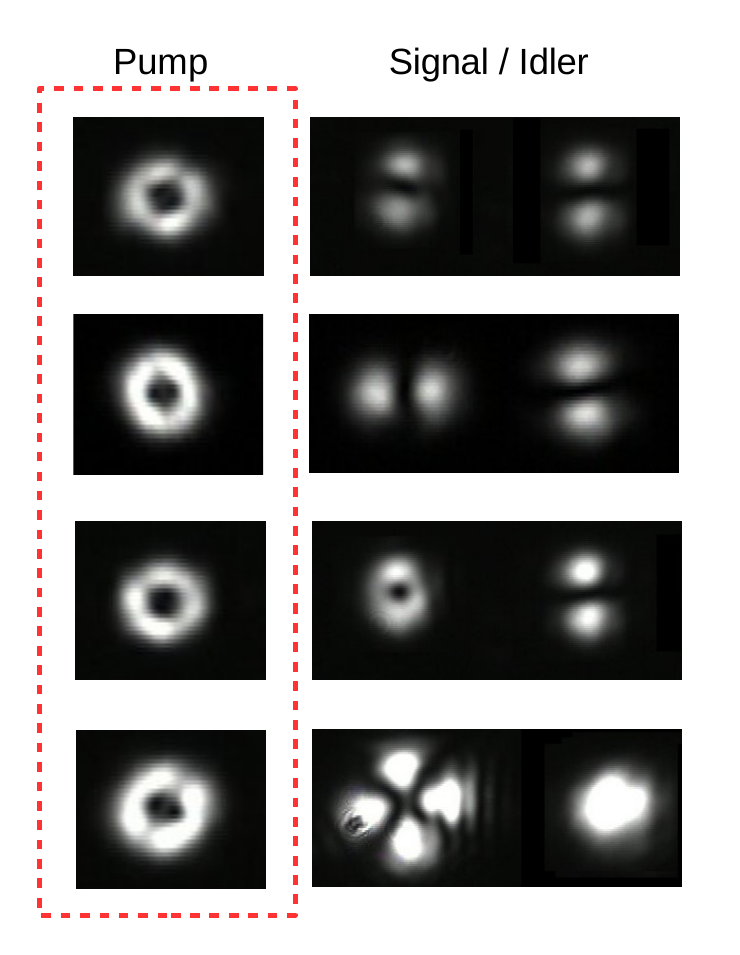}
\caption{Experimental images for the pump, signal and idler beams. The pump field is kept in the $u^{LG}_{02}$ mode.}
\label{fig-pump-02}
\end{figure}

For a pump beam prepared in a $u^{LG}_{02}$, we can figure out the possible coupling channels by inspecting its decomposition in the HG basis
\begin{equation}
u^{LG}_{02}=\frac{1}{2}\left(u^{HG}_{20}+i\sqrt{2}u^{HG}_{11}-u^{HG}_{02}\right)\,.
\end{equation}
From this decomposition we can identify the following channels
\begin{eqnarray}
u^{HG}_{20}&\rightarrow & u^{HG}_{10}+u^{HG}_{10}\,, \label{l-2-1} \\
& & u^{HG}_{20}+u^{HG}_{00}\,, \label{l-2-2} \\
\nonumber \\
u^{HG}_{11}&\rightarrow & u^{HG}_{10}+u^{HG}_{01}\,, \label{l-2-3} \\
&&u^{HG}_{11}+u^{HG}_{00}\,, \label{l-2-4} \\
\nonumber \\
u^{HG}_{02}&\rightarrow & u^{HG}_{01}+u^{HG}_{01}\,, \label{l-2-5} \\
& & u^{HG}_{02}+u^{HG}_{00} \label{l-2-6}\,.
\end{eqnarray}
Some of these possibilities are represented in Fig. \ref{fig-pump-02}, which describes one or a combination of two of the listed down-conversion processes, restricted to an idler beam in a pure Hermite-Gauss mode and order conservation. For example, the first row corresponds to process \eqref{l-2-1}; the second row, to process \eqref{l-2-3}; the third row, to a combination of  processes \eqref{l-2-1} and \eqref{l-2-3}; and finally, the fourth row, to a combination of processes \eqref{l-2-2} and \eqref{l-2-6}. 

\begin{figure}
\centering
\includegraphics[width=0.3\textwidth]{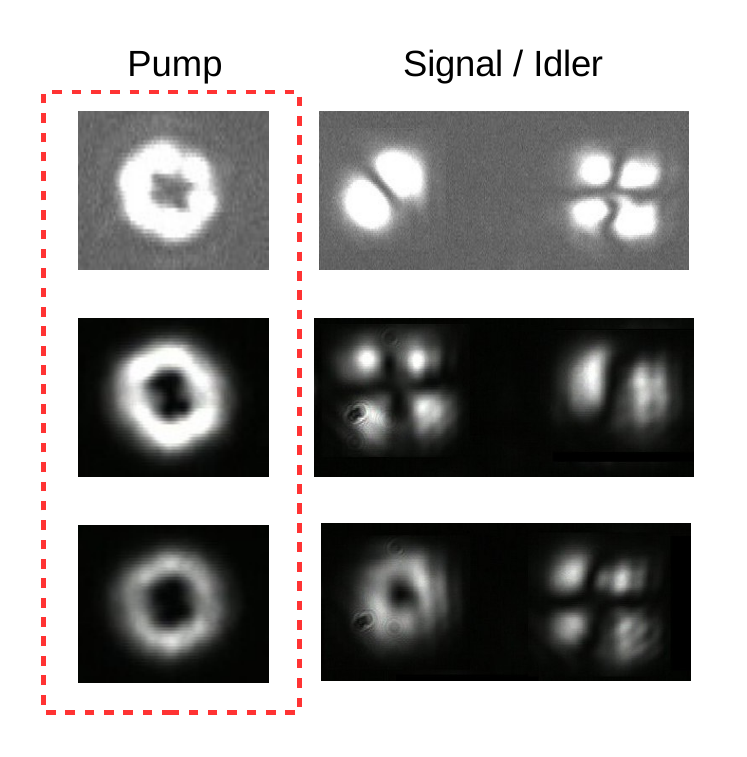}
\caption{Experimental images for the pump, signal and idler beams. The pump field is kept in the $u^{LG}_{03}$ mode.}
\label{fig-pump-03}
\end{figure}

We present in Fig.\ref{fig-pump-03} the highest OAM pump operation achieved with the laser power available in our experiment, corresponding to a $u^{LG}_{03}$ mode. Its decomposition in the HG basis writes
\begin{equation}
u^{LG}_{03}=\frac{\sqrt{2}}{4}\left(u^{HG}_{30}-\sqrt{3}u^{HG}_{12}-i\sqrt{3}u^{HG}_{21}+iu^{HG}_{03}\right)\,,
\end{equation}
giving rise to several coupling processes, with
\begin{eqnarray}
u^{HG}_{21}&\rightarrow& u^{HG}_{10}+u^{HG}_{11}\,,\label{eq-fav1} \\
u^{HG}_{12}&\rightarrow& u^{HG}_{01}+u^{HG}_{11}\,,\label{eq-fav2}
\end{eqnarray}
being the most favorable ones. The first row corresponds to a combination of processes \eqref{eq-fav1} and \eqref{eq-fav2} without OAM transfer; the second row is a pure \eqref{eq-fav2} process; and the third row is also a combination of processes \eqref{eq-fav1} and \eqref{eq-fav2} but one unit of OAM transfer (signal beam in $u^{LG}_{01}$). In all cases, total or partial OAM transfer is strongly dependent on the cavity parameters, specially the crystal orientation that affects the astigmatic anisotropy. The topological charges informed in this section were measured with the tilted lens method\cite{VAITY20131154}. The corresponding experimental results are shown in Appendix \ref{app-tilt}. 

\section{Threshold hierarchy and mode survival}\label{sec-hierarchy}

\begin{figure*}[t!]
\centering
\includegraphics[width=0.85\textwidth]{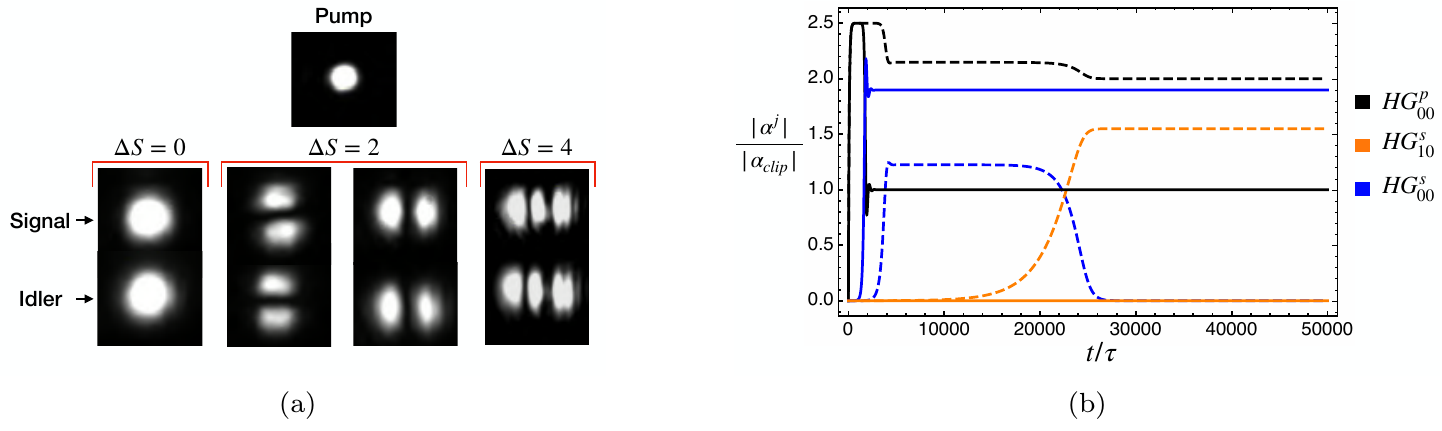}
\caption{Illustration of the mode competition and the non-conservation of the order for a $HG_{00}$ pump. In (a) we show a set of down-converted modes, with different coupling constants,  obtained at slightly different cavity lengths within a same pump resonance. In (b) we show the time evolution of the intracavity fields for the two leftmost sets of (a) for $\Delta\phi^1_{00}=\Delta\phi^2_{00}=0$ (solid lines) and $\Delta\phi^1_{00}=\Delta\phi^2_{00}=1.9\gamma '$(dashed lines), with the less coupled $HG_{10}$ in perfect resonance.}
\label{fig-nonconservation}
\end{figure*}

In the previous section, we have assigned the selection of the converted modes in the OPO to the higher coupling they have with a given pump mode, leading to the ``order conservation'' property. However, we could also observe other transverse mode operations which do not conserve the order, despite their lower coupling constants. Examples of such operation regimes can be seen in Fig. \ref{fig-nonconservation}. In that case, for slightly different cavity lengths within the same pump resonance, a Gaussian pump beam has generated multiple sets of signal and idler modes, with different coupling constants and non-zero values of $\Delta S$.

Indeed, the oscillation threshold of a given set of modes is not solely determined by the coupling constant. In particular, the mode detunings play a major role in the mode selection, since the resonance peaks of different transverse modes can be separated in an astigmatic cavity. The expression for the oscillation threshold for a three-mode coupling is given by \cite{debuisschert_type-ii_1993}
\begin{equation}
|\alpha^{in}_{mn}|^2_{th}=\frac{\gamma_0'^2\gamma_1'\gamma_2'}{2|I^{mpr}_{nqs}|^2}(1+\Delta^2)(1+\Delta_0^2)\,,
\end{equation}
where $\Delta=\delta\varphi^1_{pq}/\gamma_1'=\delta\varphi^2_{rs}/\gamma_2'$ and $\Delta_0=\delta\varphi^0_{mn}/\gamma_0'$ are the normalized detunings for the interacting modes. We then see that the normalized detunings can compensate for a smaller coupling constant in the expression for the threshold. For instance, when $(1+\Delta^2)(1+\Delta_0^2)>4$ for the $u_{00}^{HG}\rightarrow u_{00}^{HG}+u_{00}^{HG}$ conversion, its threshold becomes higher than the $u_{00}^{HG}\rightarrow u_{01}^{HG}+u_{01}^{HG}$ threshold at resonance, despite the fact that $\Lambda_{000}^{000}=2\Lambda_{011}^{000}$.

To be more rigorous and to further analyze the detuning effect on the mode selection, we have numerically integrated the dynamical equations for the OPO, including two different coupling channels with a given pump mode. As our model, we considered only the contributions of the above-mentioned channels $u_{00}^{HG}\rightarrow u_{00}^{HG}+u_{00}^{HG}$ and $u_{00}^{HG}\rightarrow u_{01}^{HG}+u_{01}^{HG}$, which are the two leftmost sets of signal and idler modes appearing in Fig \ref{fig-nonconservation}. The result is shown in Fig \ref{fig-nonconservation}, where it can be seen that a higher detuning in the stronger coupling channel may increase its threshold power, favoring the operation of the weaker coupling channel, as in the cases shown in Fig. \ref{fig-nonconservation}. This shows that the modes involved in the OPO undergo a Darwinian selection mechanism which restricts the survival to the more adapted modes. This a well known phenomenon in laser physics \cite{siegman1986lasers, haken}, but it is also present in other kinds of dynamical systems such as biological, social and economical \cite{lotka,volterra,Goel:1971}, as mentioned in \cite{sargent1974laser}.

\section{Conclusion}

In summary, we studied the transverse mode dynamics in a type-II optical parametric oscillator driven by a structured pump beam. Several effects playing a major role in the multimode dynamics were considered. First, the spatial overlap between the interacting modes was calculated, giving rise to a set of selection rules for nonvanishing intermode coupling. Then, mode selection was investigated under the influence of different aspects such as the transverse coupling strength, cavity anisotropies and mode detuning. All these effects were shown to play an important role in the oscillation threshold for different transverse mode configurations. The dynamical mode selection is determined by the configuration with lowest threshold under the coupling and cavity conditions assumed. This builds a Darwinian scenario analogous to different biological, social and economical systems, where competition allows only the most adapted element to survive.

\appendix

\section{Paraxial modes}\label{app-paraxial}

In this appendix we present the mathematical expressions and the main parameters describing the paraxial modes used in the main text. The Hermite-Gauss ($\rm{HG}_{mn}$) and the Laguerre-Gauss ($\rm{LG}_{pl}$) modes are written as

\begin{equation}\label{modo-HG}
\begin{split}
&u^{HG}_{mn}(x,y,z)= A_{mn}\frac{w_0}{w(z)}\times\\& H_m\left(\dfrac{\sqrt{2}x}{w(z)}\right) H_n\left(\dfrac{\sqrt{2}y}{w(z)}\right) \exp{\left(-\frac{x^2+y^2}{w^2(z)}\right)} \exp{\left(\phi^{HG}_{mn}(z)\right)},
\end{split}
\end{equation}
\begin{equation}\label{modo-LG}
\begin{split}
&u^{LG}_{pl}(r,\phi,z)= B_{pl}\frac{w_0}{w(z)}\times\\& \left(\frac{r}{w(z)}\right)^{|l|} L^{|l|}_p\left(\frac{2r^2}{w^2(z)}\right) \exp{\left(-\frac{r^2}{w^2(z)}\right)} \exp{\left(\phi^{LG}_{pl}(z)\right)},
\end{split}
\end{equation}
where
\begin{eqnarray}
A_{mn}&=&\sqrt{\frac{2\times 2^{-(m+n)}}{\pi w^2 m!\,n!}}\,,\\
B_{pl}&=&\sqrt{\frac{2p!}{\pi w^2(p+|l|)!}}\,,
\end{eqnarray}
are normalization constants,
\begin{eqnarray}
\phi^{HG}_{mn}(z)&=&ik\frac{x^2+y^2}{2R(z)}-i(m+n+1)\zeta(z)\,,\\
\phi^{LG}_{pl}(z)&=&\frac{ik\,r^2}{2R(z)}+il\phi-i(2p+|l|+1)\zeta(z)\,,
\end{eqnarray}
are the Gouy phase terms in the different bases, and
\begin{eqnarray}
w(z)&=&w\sqrt{1+(z/z_r)^2}\,,\\
R(z)&=&z(1+(z_r/z)^2)\,,\\
\zeta(z)&=&\tan^{-1}(z/z_r)\,,\\
w&=&\sqrt{\lambda z_r/\pi}\,.
\end{eqnarray}
In physical terms, $w$ is called the beam waist, $R(z)$ represents its wavefront curvature radius, and $z_r$ is the Rayleigh length. 

\begin{figure*}[t!]
\centering
\includegraphics[width=0.9\textwidth]{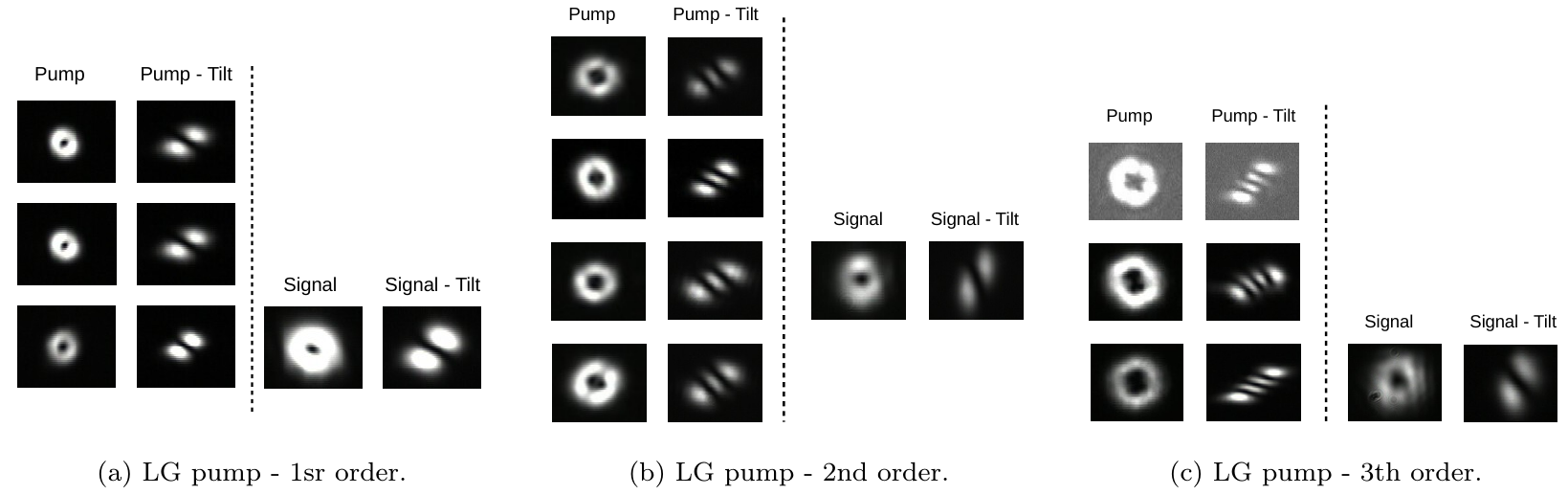}
\caption{Images for the pump and signal beams taken from Figs.\ref{fig-pump-01},\ref{fig-pump-02},\ref{fig-pump-03} and its respective conversions after the tilted lens.}
\label{fig-LG01-tilt}
\end{figure*}

\section{General expressions for the transverse coupling constants}\label{app-coupling}

Here we derive the expressions for the transverse coupling constants used in the main text. Using the generating function for the Hermite polynomials as
\begin{equation}
H_n(\gamma x)=\frac{\partial^n}{\partial t^n}\left[\exp{(2\gamma xt-t^2)}\right]\Big\vert_{t=0}\,,
\end{equation}
the integrals appearing in \eqref{eq-recobrimento-5} become
\begin{equation}
\begin{split}
&\int dx\, e^{-x^2}H_m(x)H_p(\gamma_1x)H_r(\gamma_2x)=\sqrt{\pi}\,\times\\
&\frac{\partial^m}{\partial t_0^m}\frac{\partial^p}{\partial t_1^p}\frac{\partial^r}{\partial t_2^r}\int dx\,e^{-x^2}e^{2x(t_0+\gamma_1t_1+\gamma_2t_2) - (t_0^2+t_1^2+t_2^2)}\Big\vert_{t_i=0}\,,
\end{split}
\end{equation}
which can be easily solved due to the Gaussian term in the integrand. Furthermore, since the Rayleigh length $z_r$ is fixed by the cavity geometry and signal and idler operate close to frequency degeneracy, it is reasonable to assume that they have approximately the same beam waist ($\gamma_1=\gamma_2\approx 1/\sqrt{2}$), which leads us to the following expression for the transverse overlap integral in the HG basis

\begin{equation}\label{recobrimento-HG}
\begin{split}
&\Lambda^{mpr}_{nqs}(0)=\Lambda^{000}_{000}\,C^{mpr}_{nqs}\frac{\partial^{p+r}}{\partial t_1^p\partial t_2^r}\left[(t_1+t_2)^m\,e^{-(t_1-t_2)^2/2}\right]\Big\vert_{t_i=0}\\
&\times \frac{\partial^{q+s}}{\partial t_1^q\partial t_2^s}\left[(t_1+t_2)^n\,e^{-(t_1-t_2)^2/2}\right]\Big\vert_{t_i=0}\,.
\end{split}
\end{equation}

The same procedure can be applied to the Laguerre-Gauss basis, where the functions $u^{LG}_{pl}$ in Eq.\eqref{modo-LG} are used in Eq.\eqref{eq-recobrimento}. In this case, the Laguerre polynomials can be written as
\begin{equation}
L^{|k|}_n(x)=\frac{1}{n!}\frac{\partial^n}{\partial z^n}\left[\frac{\exp{(xz/(z-1))}}{(1-z)^{|k|+1}}\right]\Big\vert_{z=0}\,,
\end{equation}
resulting in the following expression for the overlap integral
\begin{equation}\label{recobrimento-LG}
\begin{split}
&\Lambda^{lmn}_{pqr}(0)=\Lambda^{000}_{000}\,D^{lmn}_{pqr}\,2P_0!\,\delta_{l,m+n}\times\\
&\frac{\partial^{p+q+r}}{\partial z_0^p \partial z_1^q \partial z_2^r}\left[\frac{(1-z_0)^{-|l|-1}(1-z_1)^{-|m|-1}}{(1-z_2)^{|n|+1}\,\gamma^{P_0+1}}\right]\Big\vert_{z_i=0}\,,
\end{split}
\end{equation}
where
\begin{eqnarray}
D^{lmn}_{pqr}&=&\tfrac{2^{|l|/2}}{\sqrt{p!q!r!(p+|l|)!(q+|m|)!(r+|n|)!}}\,, \\
P_0&=&\tfrac{|l|+|m|+|n|}{2}\,, \\
\gamma&=& 2-\tfrac{2z_0}{z_0-1}-\tfrac{z_1}{z_1-1}-\tfrac{z_2}{z_2-1}\,.
\end{eqnarray}
Note that the conservation of angular momentum is guaranteed by the term $\delta_{l,m+n}$ \cite{Schwob1998} and the conditions (\ref{eq-conditions-1},\ref{eq-conditions-2}) still apply here, as mentioned before. The advantages of using expressions (\ref{recobrimento-HG},\ref{recobrimento-LG}) instead of their integral counterparts is that they provide a straightforward exact result that is an easy computational task.
\\
\section{Topological charge measurements}\label{app-tilt}

As mentioned in section \ref{sec-exp-setup}, the topological charge measurements of the observed LG beams were made using the technique described in \cite{VAITY20131154}. By passing a Laguerre-Gaussian beam through a tilted lens (which works as an astigmatic medium), one can convert a $u_{pl}^{LG}$ mode into a Hermite-Gauss $u^{HG}_{|l|0}$ oriented at $+(-) 45^\circ$ for positive (negative) $l$ values. More explicitly, the position where the conversion takes place depends on the distance between the original waist and the lens ($d_0$), its focal length ($f$), and the tilt angle with respect to the propagation direction ($\theta$). For a given set of parameters $(d_0, f,\theta)$, the conversion point can be found by scanning the distance after the lens directly with the CCD camera. This was done for all the Laguerre-Gauss modes present in Figs. \ref{fig-pump-01},\ref{fig-pump-02},\ref{fig-pump-03}, which are shown in Fig \ref{fig-LG01-tilt}, confirming the topological charges claimed in the main text.

\bibliographystyle{apsrev4-1} 
\bibliography{references} 

\begin{thebibliography}{42}%
\makeatletter
\providecommand \@ifxundefined [1]{%
 \@ifx{#1\undefined}
}%
\providecommand \@ifnum [1]{%
 \ifnum #1\expandafter \@firstoftwo
 \else \expandafter \@secondoftwo
 \fi
}%
\providecommand \@ifx [1]{%
 \ifx #1\expandafter \@firstoftwo
 \else \expandafter \@secondoftwo
 \fi
}%
\providecommand \natexlab [1]{#1}%
\providecommand \enquote  [1]{``#1''}%
\providecommand \bibnamefont  [1]{#1}%
\providecommand \bibfnamefont [1]{#1}%
\providecommand \citenamefont [1]{#1}%
\providecommand \href@noop [0]{\@secondoftwo}%
\providecommand \href [0]{\begingroup \@sanitize@url \@href}%
\providecommand \@href[1]{\@@startlink{#1}\@@href}%
\providecommand \@@href[1]{\endgroup#1\@@endlink}%
\providecommand \@sanitize@url [0]{\catcode `\\12\catcode `\$12\catcode
  `\&12\catcode `\#12\catcode `\^12\catcode `\_12\catcode `\%12\relax}%
\providecommand \@@startlink[1]{}%
\providecommand \@@endlink[0]{}%
\providecommand \url  [0]{\begingroup\@sanitize@url \@url }%
\providecommand \@url [1]{\endgroup\@href {#1}{\urlprefix }}%
\providecommand \urlprefix  [0]{URL }%
\providecommand \Eprint [0]{\href }%
\providecommand \doibase [0]{http://dx.doi.org/}%
\providecommand \selectlanguage [0]{\@gobble}%
\providecommand \bibinfo  [0]{\@secondoftwo}%
\providecommand \bibfield  [0]{\@secondoftwo}%
\providecommand \translation [1]{[#1]}%
\providecommand \BibitemOpen [0]{}%
\providecommand \bibitemStop [0]{}%
\providecommand \bibitemNoStop [0]{.\EOS\space}%
\providecommand \EOS [0]{\spacefactor3000\relax}%
\providecommand \BibitemShut  [1]{\csname bibitem#1\endcsname}%
\let\auto@bib@innerbib\@empty
\bibitem [{\citenamefont {Ayrapetyan}\ and\ \citenamefont
  {Fomin}(2018)}]{ayrapetyan_laser_2018}%
  \BibitemOpen
  \bibfield  {author} {\bibinfo {author} {\bibfnamefont {V.}~\bibnamefont
  {Ayrapetyan}}\ and\ \bibinfo {author} {\bibfnamefont {P.}~\bibnamefont
  {Fomin}},\ }\href {\doibase 10.1016/j.optlastec.2018.04.001} {\bibfield
  {journal} {\bibinfo  {journal} {Optics \& Laser Technology}\ }\textbf
  {\bibinfo {volume} {106}},\ \bibinfo {pages} {202} (\bibinfo {year}
  {2018})}\BibitemShut {NoStop}%
\bibitem [{\citenamefont {Vodopyanov}\ \emph {et~al.}(2018)\citenamefont
  {Vodopyanov}, \citenamefont {Muraviev}, \citenamefont {Loparo}, \citenamefont
  {Vasilyev},\ and\ \citenamefont {Mirov}}]{vodopyanov_massively_2018}%
  \BibitemOpen
  \bibfield  {author} {\bibinfo {author} {\bibfnamefont {K.~L.}\ \bibnamefont
  {Vodopyanov}}, \bibinfo {author} {\bibfnamefont {A.}~\bibnamefont
  {Muraviev}}, \bibinfo {author} {\bibfnamefont {Z.}~\bibnamefont {Loparo}},
  \bibinfo {author} {\bibfnamefont {S.}~\bibnamefont {Vasilyev}}, \ and\
  \bibinfo {author} {\bibfnamefont {S.~B.}\ \bibnamefont {Mirov}}\ }(\bibinfo
  {publisher} {SPIE},\ \bibinfo {year} {2018})\ p.\ \bibinfo {pages}
  {102}\BibitemShut {NoStop}%
\bibitem [{\citenamefont {Kusano}\ \emph {et~al.}(2018)\citenamefont {Kusano},
  \citenamefont {Hatano}, \citenamefont {Oguchi}, \citenamefont {Yamawaki},
  \citenamefont {Watanabe},\ and\ \citenamefont {Enoki}}]{kusano_mid-ir_2018}%
  \BibitemOpen
  \bibfield  {author} {\bibinfo {author} {\bibfnamefont {M.}~\bibnamefont
  {Kusano}}, \bibinfo {author} {\bibfnamefont {H.}~\bibnamefont {Hatano}},
  \bibinfo {author} {\bibfnamefont {K.}~\bibnamefont {Oguchi}}, \bibinfo
  {author} {\bibfnamefont {H.}~\bibnamefont {Yamawaki}}, \bibinfo {author}
  {\bibfnamefont {M.}~\bibnamefont {Watanabe}}, \ and\ \bibinfo {author}
  {\bibfnamefont {M.}~\bibnamefont {Enoki}}\ }(\bibinfo {year} {2018})\ p.\
  \bibinfo {pages} {180005}\BibitemShut {NoStop}%
\bibitem [{\citenamefont {Kaskow}\ \emph {et~al.}(2018)\citenamefont {Kaskow},
  \citenamefont {Gorajek}, \citenamefont {Zendzian},\ and\ \citenamefont
  {Jabczynski}}]{kaskow_mw_2018}%
  \BibitemOpen
  \bibfield  {author} {\bibinfo {author} {\bibfnamefont {M.}~\bibnamefont
  {Kaskow}}, \bibinfo {author} {\bibfnamefont {L.}~\bibnamefont {Gorajek}},
  \bibinfo {author} {\bibfnamefont {W.}~\bibnamefont {Zendzian}}, \ and\
  \bibinfo {author} {\bibfnamefont {J.}~\bibnamefont {Jabczynski}},\ }\href
  {\doibase 10.1016/j.opelre.2018.04.005} {\bibfield  {journal} {\bibinfo
  {journal} {Opto-Electronics Review}\ }\textbf {\bibinfo {volume} {26}},\
  \bibinfo {pages} {188} (\bibinfo {year} {2018})}\BibitemShut {NoStop}%
\bibitem [{\citenamefont {Heidmann}\ \emph {et~al.}(1987)\citenamefont
  {Heidmann}, \citenamefont {Horowicz}, \citenamefont {Reynaud}, \citenamefont
  {Giacobino}, \citenamefont {Fabre},\ and\ \citenamefont
  {Camy}}]{PhysRevLett.59.2555}%
  \BibitemOpen
  \bibfield  {author} {\bibinfo {author} {\bibfnamefont {A.}~\bibnamefont
  {Heidmann}}, \bibinfo {author} {\bibfnamefont {R.~J.}\ \bibnamefont
  {Horowicz}}, \bibinfo {author} {\bibfnamefont {S.}~\bibnamefont {Reynaud}},
  \bibinfo {author} {\bibfnamefont {E.}~\bibnamefont {Giacobino}}, \bibinfo
  {author} {\bibfnamefont {C.}~\bibnamefont {Fabre}}, \ and\ \bibinfo {author}
  {\bibfnamefont {G.}~\bibnamefont {Camy}},\ }\href {\doibase
  10.1103/PhysRevLett.59.2555} {\bibfield  {journal} {\bibinfo  {journal}
  {Phys. Rev. Lett.}\ }\textbf {\bibinfo {volume} {59}},\ \bibinfo {pages}
  {2555} (\bibinfo {year} {1987})}\BibitemShut {NoStop}%
\bibitem [{\citenamefont {Mertz}\ \emph {et~al.}(1991)\citenamefont {Mertz},
  \citenamefont {Debuisschert}, \citenamefont {Heidmann}, \citenamefont
  {Fabre},\ and\ \citenamefont {Giacobino}}]{Mertz:91}%
  \BibitemOpen
  \bibfield  {author} {\bibinfo {author} {\bibfnamefont {J.}~\bibnamefont
  {Mertz}}, \bibinfo {author} {\bibfnamefont {T.}~\bibnamefont {Debuisschert}},
  \bibinfo {author} {\bibfnamefont {A.}~\bibnamefont {Heidmann}}, \bibinfo
  {author} {\bibfnamefont {C.}~\bibnamefont {Fabre}}, \ and\ \bibinfo {author}
  {\bibfnamefont {E.}~\bibnamefont {Giacobino}},\ }\href {\doibase
  10.1364/OL.16.001234} {\bibfield  {journal} {\bibinfo  {journal} {Opt.
  Lett.}\ }\textbf {\bibinfo {volume} {16}},\ \bibinfo {pages} {1234} (\bibinfo
  {year} {1991})}\BibitemShut {NoStop}%
\bibitem [{\citenamefont {Ou}\ \emph {et~al.}(1992)\citenamefont {Ou},
  \citenamefont {Pereira}, \citenamefont {Kimble},\ and\ \citenamefont
  {Peng}}]{PhysRevLett.68.3663}%
  \BibitemOpen
  \bibfield  {author} {\bibinfo {author} {\bibfnamefont {Z.~Y.}\ \bibnamefont
  {Ou}}, \bibinfo {author} {\bibfnamefont {S.~F.}\ \bibnamefont {Pereira}},
  \bibinfo {author} {\bibfnamefont {H.~J.}\ \bibnamefont {Kimble}}, \ and\
  \bibinfo {author} {\bibfnamefont {K.~C.}\ \bibnamefont {Peng}},\ }\href
  {\doibase 10.1103/PhysRevLett.68.3663} {\bibfield  {journal} {\bibinfo
  {journal} {Phys. Rev. Lett.}\ }\textbf {\bibinfo {volume} {68}},\ \bibinfo
  {pages} {3663} (\bibinfo {year} {1992})}\BibitemShut {NoStop}%
\bibitem [{\citenamefont {Keller}\ \emph {et~al.}(2008)\citenamefont {Keller},
  \citenamefont {D'Auria}, \citenamefont {Treps}, \citenamefont {Coudreau},
  \citenamefont {Laurat},\ and\ \citenamefont {Fabre}}]{Keller:08}%
  \BibitemOpen
  \bibfield  {author} {\bibinfo {author} {\bibfnamefont {G.}~\bibnamefont
  {Keller}}, \bibinfo {author} {\bibfnamefont {V.}~\bibnamefont {D'Auria}},
  \bibinfo {author} {\bibfnamefont {N.}~\bibnamefont {Treps}}, \bibinfo
  {author} {\bibfnamefont {T.}~\bibnamefont {Coudreau}}, \bibinfo {author}
  {\bibfnamefont {J.}~\bibnamefont {Laurat}}, \ and\ \bibinfo {author}
  {\bibfnamefont {C.}~\bibnamefont {Fabre}},\ }\href {\doibase
  10.1364/OE.16.009351} {\bibfield  {journal} {\bibinfo  {journal} {Opt.
  Express}\ }\textbf {\bibinfo {volume} {16}},\ \bibinfo {pages} {9351}
  (\bibinfo {year} {2008})}\BibitemShut {NoStop}%
\bibitem [{\citenamefont {Braunstein}\ and\ \citenamefont
  {Pati}(2012)}]{braunstein2012quantum}%
  \BibitemOpen
  \bibfield  {author} {\bibinfo {author} {\bibfnamefont {S.}~\bibnamefont
  {Braunstein}}\ and\ \bibinfo {author} {\bibfnamefont {A.}~\bibnamefont
  {Pati}},\ }\href {https://books.google.com.br/books?id=PTfpCAAAQBAJ} {\emph
  {\bibinfo {title} {Quantum Information with Continuous Variables}}}\
  (\bibinfo  {publisher} {Springer Netherlands},\ \bibinfo {year}
  {2012})\BibitemShut {NoStop}%
\bibitem [{\citenamefont {Chen}\ \emph {et~al.}(2014)\citenamefont {Chen},
  \citenamefont {Menicucci},\ and\ \citenamefont
  {Pfister}}]{PhysRevLett.112.120505}%
  \BibitemOpen
  \bibfield  {author} {\bibinfo {author} {\bibfnamefont {M.}~\bibnamefont
  {Chen}}, \bibinfo {author} {\bibfnamefont {N.~C.}\ \bibnamefont {Menicucci}},
  \ and\ \bibinfo {author} {\bibfnamefont {O.}~\bibnamefont {Pfister}},\ }\href
  {\doibase 10.1103/PhysRevLett.112.120505} {\bibfield  {journal} {\bibinfo
  {journal} {Phys. Rev. Lett.}\ }\textbf {\bibinfo {volume} {112}},\ \bibinfo
  {pages} {120505} (\bibinfo {year} {2014})}\BibitemShut {NoStop}%
\bibitem [{\citenamefont {Pysher}\ \emph {et~al.}(2011)\citenamefont {Pysher},
  \citenamefont {Miwa}, \citenamefont {Shahrokhshahi}, \citenamefont
  {Bloomer},\ and\ \citenamefont {Pfister}}]{PhysRevLett.107.030505}%
  \BibitemOpen
  \bibfield  {author} {\bibinfo {author} {\bibfnamefont {M.}~\bibnamefont
  {Pysher}}, \bibinfo {author} {\bibfnamefont {Y.}~\bibnamefont {Miwa}},
  \bibinfo {author} {\bibfnamefont {R.}~\bibnamefont {Shahrokhshahi}}, \bibinfo
  {author} {\bibfnamefont {R.}~\bibnamefont {Bloomer}}, \ and\ \bibinfo
  {author} {\bibfnamefont {O.}~\bibnamefont {Pfister}},\ }\href {\doibase
  10.1103/PhysRevLett.107.030505} {\bibfield  {journal} {\bibinfo  {journal}
  {Phys. Rev. Lett.}\ }\textbf {\bibinfo {volume} {107}},\ \bibinfo {pages}
  {030505} (\bibinfo {year} {2011})}\BibitemShut {NoStop}%
\bibitem [{\citenamefont {Souza}\ \emph {et~al.}(2008)\citenamefont {Souza},
  \citenamefont {Borges}, \citenamefont {Khoury}, \citenamefont {Huguenin},
  \citenamefont {Aolita},\ and\ \citenamefont {Walborn}}]{PhysRevA.77.032345}%
  \BibitemOpen
  \bibfield  {author} {\bibinfo {author} {\bibfnamefont {C.~E.~R.}\
  \bibnamefont {Souza}}, \bibinfo {author} {\bibfnamefont {C.~V.~S.}\
  \bibnamefont {Borges}}, \bibinfo {author} {\bibfnamefont {A.~Z.}\
  \bibnamefont {Khoury}}, \bibinfo {author} {\bibfnamefont {J.~A.~O.}\
  \bibnamefont {Huguenin}}, \bibinfo {author} {\bibfnamefont {L.}~\bibnamefont
  {Aolita}}, \ and\ \bibinfo {author} {\bibfnamefont {S.~P.}\ \bibnamefont
  {Walborn}},\ }\href {\doibase 10.1103/PhysRevA.77.032345} {\bibfield
  {journal} {\bibinfo  {journal} {Phys. Rev. A}\ }\textbf {\bibinfo {volume}
  {77}},\ \bibinfo {pages} {032345} (\bibinfo {year} {2008})}\BibitemShut
  {NoStop}%
\bibitem [{\citenamefont {D'Ambrosio}\ \emph {et~al.}(2012)\citenamefont
  {D'Ambrosio}, \citenamefont {Nagali}, \citenamefont {Walborn}, \citenamefont
  {Aolita}, \citenamefont {Slussarenko}, \citenamefont {Marrucci},\ and\
  \citenamefont {Sciarrino}}]{dambrosio_complete_2012}%
  \BibitemOpen
  \bibfield  {author} {\bibinfo {author} {\bibfnamefont {V.}~\bibnamefont
  {D'Ambrosio}}, \bibinfo {author} {\bibfnamefont {E.}~\bibnamefont {Nagali}},
  \bibinfo {author} {\bibfnamefont {S.~P.}\ \bibnamefont {Walborn}}, \bibinfo
  {author} {\bibfnamefont {L.}~\bibnamefont {Aolita}}, \bibinfo {author}
  {\bibfnamefont {S.}~\bibnamefont {Slussarenko}}, \bibinfo {author}
  {\bibfnamefont {L.}~\bibnamefont {Marrucci}}, \ and\ \bibinfo {author}
  {\bibfnamefont {F.}~\bibnamefont {Sciarrino}},\ }\href {\doibase
  10.1038/ncomms1951} {\bibfield  {journal} {\bibinfo  {journal} {Nature
  Communications}\ }\textbf {\bibinfo {volume} {3}} (\bibinfo {year} {2012}),\
  10.1038/ncomms1951}\BibitemShut {NoStop}%
\bibitem [{\citenamefont {Mair}\ \emph {et~al.}(2001)\citenamefont {Mair},
  \citenamefont {Vaziri}, \citenamefont {Weihs},\ and\ \citenamefont
  {Zeilinger}}]{mair_entanglement_2001}%
  \BibitemOpen
  \bibfield  {author} {\bibinfo {author} {\bibfnamefont {A.}~\bibnamefont
  {Mair}}, \bibinfo {author} {\bibfnamefont {A.}~\bibnamefont {Vaziri}},
  \bibinfo {author} {\bibfnamefont {G.}~\bibnamefont {Weihs}}, \ and\ \bibinfo
  {author} {\bibfnamefont {A.}~\bibnamefont {Zeilinger}},\ }\href {\doibase
  10.1038/35085529} {\bibfield  {journal} {\bibinfo  {journal} {Nature}\
  }\textbf {\bibinfo {volume} {412}},\ \bibinfo {pages} {313} (\bibinfo {year}
  {2001})}\BibitemShut {NoStop}%
\bibitem [{\citenamefont {Lassen}\ \emph {et~al.}(2009)\citenamefont {Lassen},
  \citenamefont {Leuchs},\ and\ \citenamefont
  {Andersen}}]{PhysRevLett.102.163602}%
  \BibitemOpen
  \bibfield  {author} {\bibinfo {author} {\bibfnamefont {M.}~\bibnamefont
  {Lassen}}, \bibinfo {author} {\bibfnamefont {G.}~\bibnamefont {Leuchs}}, \
  and\ \bibinfo {author} {\bibfnamefont {U.~L.}\ \bibnamefont {Andersen}},\
  }\href {\doibase 10.1103/PhysRevLett.102.163602} {\bibfield  {journal}
  {\bibinfo  {journal} {Phys. Rev. Lett.}\ }\textbf {\bibinfo {volume} {102}},\
  \bibinfo {pages} {163602} (\bibinfo {year} {2009})}\BibitemShut {NoStop}%
\bibitem [{\citenamefont {Gahagan}\ and\ \citenamefont
  {Swartzlander}(1996)}]{Gahagan:96}%
  \BibitemOpen
  \bibfield  {author} {\bibinfo {author} {\bibfnamefont {K.~T.}\ \bibnamefont
  {Gahagan}}\ and\ \bibinfo {author} {\bibfnamefont {G.~A.}\ \bibnamefont
  {Swartzlander}},\ }\href {\doibase 10.1364/OL.21.000827} {\bibfield
  {journal} {\bibinfo  {journal} {Opt. Lett.}\ }\textbf {\bibinfo {volume}
  {21}},\ \bibinfo {pages} {827} (\bibinfo {year} {1996})}\BibitemShut
  {NoStop}%
\bibitem [{\citenamefont {Foo}\ \emph {et~al.}(2005)\citenamefont {Foo},
  \citenamefont {Palacios},\ and\ \citenamefont {Swartzlander}}]{Foo:05}%
  \BibitemOpen
  \bibfield  {author} {\bibinfo {author} {\bibfnamefont {G.}~\bibnamefont
  {Foo}}, \bibinfo {author} {\bibfnamefont {D.~M.}\ \bibnamefont {Palacios}}, \
  and\ \bibinfo {author} {\bibfnamefont {G.~A.}\ \bibnamefont {Swartzlander}},\
  }\href {\doibase 10.1364/OL.30.003308} {\bibfield  {journal} {\bibinfo
  {journal} {Opt. Lett.}\ }\textbf {\bibinfo {volume} {30}},\ \bibinfo {pages}
  {3308} (\bibinfo {year} {2005})}\BibitemShut {NoStop}%
\bibitem [{\citenamefont {Bramati}\ \emph {et~al.}(1999)\citenamefont
  {Bramati}, \citenamefont {Hermier}, \citenamefont {Khoury}, \citenamefont
  {Giacobino}, \citenamefont {Schnitzer}, \citenamefont {Michalzik},
  \citenamefont {Ebeling}, \citenamefont {Poizat},\ and\ \citenamefont
  {Grangier}}]{Bramati:99}%
  \BibitemOpen
  \bibfield  {author} {\bibinfo {author} {\bibfnamefont {A.}~\bibnamefont
  {Bramati}}, \bibinfo {author} {\bibfnamefont {J.-P.}\ \bibnamefont
  {Hermier}}, \bibinfo {author} {\bibfnamefont {A.~Z.}\ \bibnamefont {Khoury}},
  \bibinfo {author} {\bibfnamefont {E.}~\bibnamefont {Giacobino}}, \bibinfo
  {author} {\bibfnamefont {P.}~\bibnamefont {Schnitzer}}, \bibinfo {author}
  {\bibfnamefont {R.}~\bibnamefont {Michalzik}}, \bibinfo {author}
  {\bibfnamefont {K.~J.}\ \bibnamefont {Ebeling}}, \bibinfo {author}
  {\bibfnamefont {J.-P.}\ \bibnamefont {Poizat}}, \ and\ \bibinfo {author}
  {\bibfnamefont {P.}~\bibnamefont {Grangier}},\ }\href {\doibase
  10.1364/OL.24.000893} {\bibfield  {journal} {\bibinfo  {journal} {Opt.
  Lett.}\ }\textbf {\bibinfo {volume} {24}},\ \bibinfo {pages} {893} (\bibinfo
  {year} {1999})}\BibitemShut {NoStop}%
\bibitem [{\citenamefont {Hermier}\ \emph {et~al.}(1999)\citenamefont
  {Hermier}, \citenamefont {Bramati}, \citenamefont {Khoury}, \citenamefont
  {Giacobino}, \citenamefont {Poizat}, \citenamefont {Chang},\ and\
  \citenamefont {Grangier}}]{Hermier:99}%
  \BibitemOpen
  \bibfield  {author} {\bibinfo {author} {\bibfnamefont {J.-P.}\ \bibnamefont
  {Hermier}}, \bibinfo {author} {\bibfnamefont {A.}~\bibnamefont {Bramati}},
  \bibinfo {author} {\bibfnamefont {A.~Z.}\ \bibnamefont {Khoury}}, \bibinfo
  {author} {\bibfnamefont {E.}~\bibnamefont {Giacobino}}, \bibinfo {author}
  {\bibfnamefont {J.-P.}\ \bibnamefont {Poizat}}, \bibinfo {author}
  {\bibfnamefont {T.~J.}\ \bibnamefont {Chang}}, \ and\ \bibinfo {author}
  {\bibfnamefont {P.}~\bibnamefont {Grangier}},\ }\href {\doibase
  10.1364/JOSAB.16.002140} {\bibfield  {journal} {\bibinfo  {journal} {J. Opt.
  Soc. Am. B}\ }\textbf {\bibinfo {volume} {16}},\ \bibinfo {pages} {2140}
  (\bibinfo {year} {1999})}\BibitemShut {NoStop}%
\bibitem [{\citenamefont {Hermier}\ \emph {et~al.}(2001)\citenamefont
  {Hermier}, \citenamefont {Bramati}, \citenamefont {Khoury}, \citenamefont
  {Josse}, \citenamefont {Giacobino}, \citenamefont {Schnitzer}, \citenamefont
  {Michalzik},\ and\ \citenamefont {Ebeling}}]{892729}%
  \BibitemOpen
  \bibfield  {author} {\bibinfo {author} {\bibfnamefont {J.~.}\ \bibnamefont
  {Hermier}}, \bibinfo {author} {\bibfnamefont {A.}~\bibnamefont {Bramati}},
  \bibinfo {author} {\bibfnamefont {A.~Z.}\ \bibnamefont {Khoury}}, \bibinfo
  {author} {\bibfnamefont {V.}~\bibnamefont {Josse}}, \bibinfo {author}
  {\bibfnamefont {E.}~\bibnamefont {Giacobino}}, \bibinfo {author}
  {\bibfnamefont {P.}~\bibnamefont {Schnitzer}}, \bibinfo {author}
  {\bibfnamefont {R.}~\bibnamefont {Michalzik}}, \ and\ \bibinfo {author}
  {\bibfnamefont {K.~J.}\ \bibnamefont {Ebeling}},\ }\href {\doibase
  10.1109/3.892729} {\bibfield  {journal} {\bibinfo  {journal} {IEEE Journal of
  Quantum Electronics}\ }\textbf {\bibinfo {volume} {37}},\ \bibinfo {pages}
  {87} (\bibinfo {year} {2001})}\BibitemShut {NoStop}%
\bibitem [{\citenamefont {Oppo}\ \emph {et~al.}(1994)\citenamefont {Oppo},
  \citenamefont {Brambilla},\ and\ \citenamefont {Lugiato}}]{PhysRevA.49.2028}%
  \BibitemOpen
  \bibfield  {author} {\bibinfo {author} {\bibfnamefont {G.-L.}\ \bibnamefont
  {Oppo}}, \bibinfo {author} {\bibfnamefont {M.}~\bibnamefont {Brambilla}}, \
  and\ \bibinfo {author} {\bibfnamefont {L.~A.}\ \bibnamefont {Lugiato}},\
  }\href {\doibase 10.1103/PhysRevA.49.2028} {\bibfield  {journal} {\bibinfo
  {journal} {Phys. Rev. A}\ }\textbf {\bibinfo {volume} {49}},\ \bibinfo
  {pages} {2028} (\bibinfo {year} {1994})}\BibitemShut {NoStop}%
\bibitem [{\citenamefont {Marte}\ \emph {et~al.}(1998)\citenamefont {Marte},
  \citenamefont {Ritsch}, \citenamefont {Petsas}, \citenamefont {Gatti},
  \citenamefont {Lugiato}, \citenamefont {Fabre},\ and\ \citenamefont
  {Leduc}}]{Marte:98}%
  \BibitemOpen
  \bibfield  {author} {\bibinfo {author} {\bibfnamefont {M.}~\bibnamefont
  {Marte}}, \bibinfo {author} {\bibfnamefont {H.}~\bibnamefont {Ritsch}},
  \bibinfo {author} {\bibfnamefont {K.~I.}\ \bibnamefont {Petsas}}, \bibinfo
  {author} {\bibfnamefont {A.}~\bibnamefont {Gatti}}, \bibinfo {author}
  {\bibfnamefont {L.~A.}\ \bibnamefont {Lugiato}}, \bibinfo {author}
  {\bibfnamefont {C.}~\bibnamefont {Fabre}}, \ and\ \bibinfo {author}
  {\bibfnamefont {D.}~\bibnamefont {Leduc}},\ }\href {\doibase
  10.1364/OE.3.000476} {\bibfield  {journal} {\bibinfo  {journal} {Opt.
  Express}\ }\textbf {\bibinfo {volume} {3}},\ \bibinfo {pages} {476} (\bibinfo
  {year} {1998})}\BibitemShut {NoStop}%
\bibitem [{\citenamefont {Vaupel}\ \emph {et~al.}(1999)\citenamefont {Vaupel},
  \citenamefont {Ma\^{\i}tre},\ and\ \citenamefont
  {Fabre}}]{PhysRevLett.83.5278}%
  \BibitemOpen
  \bibfield  {author} {\bibinfo {author} {\bibfnamefont {M.}~\bibnamefont
  {Vaupel}}, \bibinfo {author} {\bibfnamefont {A.}~\bibnamefont {Ma\^{\i}tre}},
  \ and\ \bibinfo {author} {\bibfnamefont {C.}~\bibnamefont {Fabre}},\ }\href
  {\doibase 10.1103/PhysRevLett.83.5278} {\bibfield  {journal} {\bibinfo
  {journal} {Phys. Rev. Lett.}\ }\textbf {\bibinfo {volume} {83}},\ \bibinfo
  {pages} {5278} (\bibinfo {year} {1999})}\BibitemShut {NoStop}%
\bibitem [{\citenamefont {Ducci}\ \emph {et~al.}(2001)\citenamefont {Ducci},
  \citenamefont {Treps}, \citenamefont {Ma\^{\i}tre},\ and\ \citenamefont
  {Fabre}}]{PhysRevA.64.023803}%
  \BibitemOpen
  \bibfield  {author} {\bibinfo {author} {\bibfnamefont {S.}~\bibnamefont
  {Ducci}}, \bibinfo {author} {\bibfnamefont {N.}~\bibnamefont {Treps}},
  \bibinfo {author} {\bibfnamefont {A.}~\bibnamefont {Ma\^{\i}tre}}, \ and\
  \bibinfo {author} {\bibfnamefont {C.}~\bibnamefont {Fabre}},\ }\href
  {\doibase 10.1103/PhysRevA.64.023803} {\bibfield  {journal} {\bibinfo
  {journal} {Phys. Rev. A}\ }\textbf {\bibinfo {volume} {64}},\ \bibinfo
  {pages} {023803} (\bibinfo {year} {2001})}\BibitemShut {NoStop}%
\bibitem [{\citenamefont {Schwob}\ \emph {et~al.}(1998)\citenamefont {Schwob},
  \citenamefont {Cohadon}, \citenamefont {Fabre}, \citenamefont {Marte},
  \citenamefont {Ritsch}, \citenamefont {Gatti},\ and\ \citenamefont
  {Lugiato}}]{Schwob1998}%
  \BibitemOpen
  \bibfield  {author} {\bibinfo {author} {\bibfnamefont {C.}~\bibnamefont
  {Schwob}}, \bibinfo {author} {\bibfnamefont {P.}~\bibnamefont {Cohadon}},
  \bibinfo {author} {\bibfnamefont {C.}~\bibnamefont {Fabre}}, \bibinfo
  {author} {\bibfnamefont {M.}~\bibnamefont {Marte}}, \bibinfo {author}
  {\bibfnamefont {H.}~\bibnamefont {Ritsch}}, \bibinfo {author} {\bibfnamefont
  {A.}~\bibnamefont {Gatti}}, \ and\ \bibinfo {author} {\bibfnamefont
  {L.}~\bibnamefont {Lugiato}},\ }\href {\doibase 10.1007/s003400050455}
  {\bibfield  {journal} {\bibinfo  {journal} {Applied Physics B}\ }\textbf
  {\bibinfo {volume} {66}},\ \bibinfo {pages} {685} (\bibinfo {year}
  {1998})}\BibitemShut {NoStop}%
\bibitem [{\citenamefont {Martinelli}\ \emph {et~al.}(2004)\citenamefont
  {Martinelli}, \citenamefont {Huguenin}, \citenamefont {Nussenzveig},\ and\
  \citenamefont {Khoury}}]{PhysRevA.70.013812}%
  \BibitemOpen
  \bibfield  {author} {\bibinfo {author} {\bibfnamefont {M.}~\bibnamefont
  {Martinelli}}, \bibinfo {author} {\bibfnamefont {J.~A.~O.}\ \bibnamefont
  {Huguenin}}, \bibinfo {author} {\bibfnamefont {P.}~\bibnamefont
  {Nussenzveig}}, \ and\ \bibinfo {author} {\bibfnamefont {A.~Z.}\ \bibnamefont
  {Khoury}},\ }\href {\doibase 10.1103/PhysRevA.70.013812} {\bibfield
  {journal} {\bibinfo  {journal} {Phys. Rev. A}\ }\textbf {\bibinfo {volume}
  {70}},\ \bibinfo {pages} {013812} (\bibinfo {year} {2004})}\BibitemShut
  {NoStop}%
\bibitem [{\citenamefont {Walborn}\ and\ \citenamefont
  {Pimentel}(2012)}]{Walborn:2012}%
  \BibitemOpen
  \bibfield  {author} {\bibinfo {author} {\bibfnamefont {S.~P.}\ \bibnamefont
  {Walborn}}\ and\ \bibinfo {author} {\bibfnamefont {A.~H.}\ \bibnamefont
  {Pimentel}},\ }\href {https://doi.org/10.1088/0953-4075/45/16/165502}
  {\bibfield  {journal} {\bibinfo  {journal} {J. Phys. B: At. Mol. Opt. Phys.}\
  }\textbf {\bibinfo {volume} {45}} (\bibinfo {year} {2012})}\BibitemShut
  {NoStop}%
\bibitem [{\citenamefont {dos Santos}\ \emph {et~al.}(2008)\citenamefont {dos
  Santos}, \citenamefont {Khoury},\ and\ \citenamefont
  {Huguenin}}]{dosSantos:08}%
  \BibitemOpen
  \bibfield  {author} {\bibinfo {author} {\bibfnamefont {B.~C.}\ \bibnamefont
  {dos Santos}}, \bibinfo {author} {\bibfnamefont {A.~Z.}\ \bibnamefont
  {Khoury}}, \ and\ \bibinfo {author} {\bibfnamefont {J.~A.~O.}\ \bibnamefont
  {Huguenin}},\ }\href {\doibase 10.1364/OL.33.002803} {\bibfield  {journal}
  {\bibinfo  {journal} {Opt. Lett.}\ }\textbf {\bibinfo {volume} {33}},\
  \bibinfo {pages} {2803} (\bibinfo {year} {2008})}\BibitemShut {NoStop}%
\bibitem [{\citenamefont {Aadhi}\ \emph {et~al.}(2017)\citenamefont {Aadhi},
  \citenamefont {Samanta}, \citenamefont {Kumar},\ and\ \citenamefont
  {Ebrahim-Zadeh}}]{Aadhi:17}%
  \BibitemOpen
  \bibfield  {author} {\bibinfo {author} {\bibfnamefont {A.}~\bibnamefont
  {Aadhi}}, \bibinfo {author} {\bibfnamefont {G.~K.}\ \bibnamefont {Samanta}},
  \bibinfo {author} {\bibfnamefont {S.~C.}\ \bibnamefont {Kumar}}, \ and\
  \bibinfo {author} {\bibfnamefont {M.}~\bibnamefont {Ebrahim-Zadeh}},\ }\href
  {\doibase 10.1364/OPTICA.4.000349} {\bibfield  {journal} {\bibinfo  {journal}
  {Optica}\ }\textbf {\bibinfo {volume} {4}},\ \bibinfo {pages} {349} (\bibinfo
  {year} {2017})}\BibitemShut {NoStop}%
\bibitem [{\citenamefont {Boyd}\ and\ \citenamefont
  {Prato}(2008)}]{boyd2008nonlinear}%
  \BibitemOpen
  \bibfield  {author} {\bibinfo {author} {\bibfnamefont {R.}~\bibnamefont
  {Boyd}}\ and\ \bibinfo {author} {\bibfnamefont {D.}~\bibnamefont {Prato}},\
  }\href {https://books.google.com.br/books?id=uoRUi1Yb7ooC} {\emph {\bibinfo
  {title} {Nonlinear Optics}}},\ Nonlinear Optics Series\ (\bibinfo
  {publisher} {Elsevier Science},\ \bibinfo {year} {2008})\BibitemShut
  {NoStop}%
\bibitem [{\citenamefont {Kogelnik}\ and\ \citenamefont
  {Li}(1966)}]{Kogelnik:66}%
  \BibitemOpen
  \bibfield  {author} {\bibinfo {author} {\bibfnamefont {H.}~\bibnamefont
  {Kogelnik}}\ and\ \bibinfo {author} {\bibfnamefont {T.}~\bibnamefont {Li}},\
  }\href {\doibase 10.1364/AO.5.001550} {\bibfield  {journal} {\bibinfo
  {journal} {Appl. Opt.}\ }\textbf {\bibinfo {volume} {5}},\ \bibinfo {pages}
  {1550} (\bibinfo {year} {1966})}\BibitemShut {NoStop}%
\bibitem [{\citenamefont {Pereira}\ \emph {et~al.}(2017)\citenamefont
  {Pereira}, \citenamefont {Buono}, \citenamefont {Tasca}, \citenamefont
  {Dechoum},\ and\ \citenamefont {Khoury}}]{PhysRevA.96.053856}%
  \BibitemOpen
  \bibfield  {author} {\bibinfo {author} {\bibfnamefont {L.~J.}\ \bibnamefont
  {Pereira}}, \bibinfo {author} {\bibfnamefont {W.~T.}\ \bibnamefont {Buono}},
  \bibinfo {author} {\bibfnamefont {D.~S.}\ \bibnamefont {Tasca}}, \bibinfo
  {author} {\bibfnamefont {K.}~\bibnamefont {Dechoum}}, \ and\ \bibinfo
  {author} {\bibfnamefont {A.~Z.}\ \bibnamefont {Khoury}},\ }\href {\doibase
  10.1103/PhysRevA.96.053856} {\bibfield  {journal} {\bibinfo  {journal} {Phys.
  Rev. A}\ }\textbf {\bibinfo {volume} {96}},\ \bibinfo {pages} {053856}
  (\bibinfo {year} {2017})}\BibitemShut {NoStop}%
\bibitem [{\citenamefont {Buono}\ \emph {et~al.}(2018)\citenamefont {Buono},
  \citenamefont {Santiago}, \citenamefont {Pereira}, \citenamefont {Tasca},
  \citenamefont {Dechoum},\ and\ \citenamefont {Khoury}}]{Buono:18}%
  \BibitemOpen
  \bibfield  {author} {\bibinfo {author} {\bibfnamefont {W.~T.}\ \bibnamefont
  {Buono}}, \bibinfo {author} {\bibfnamefont {J.}~\bibnamefont {Santiago}},
  \bibinfo {author} {\bibfnamefont {L.~J.}\ \bibnamefont {Pereira}}, \bibinfo
  {author} {\bibfnamefont {D.~S.}\ \bibnamefont {Tasca}}, \bibinfo {author}
  {\bibfnamefont {K.}~\bibnamefont {Dechoum}}, \ and\ \bibinfo {author}
  {\bibfnamefont {A.~Z.}\ \bibnamefont {Khoury}},\ }\href {\doibase
  10.1364/OL.43.001439} {\bibfield  {journal} {\bibinfo  {journal} {Opt.
  Lett.}\ }\textbf {\bibinfo {volume} {43}},\ \bibinfo {pages} {1439} (\bibinfo
  {year} {2018})}\BibitemShut {NoStop}%
\bibitem [{\citenamefont {Clark}\ \emph {et~al.}(2016)\citenamefont {Clark},
  \citenamefont {Offer}, \citenamefont {Franke-Arnold}, \citenamefont
  {Arnold},\ and\ \citenamefont {Radwell}}]{Clark:16}%
  \BibitemOpen
  \bibfield  {author} {\bibinfo {author} {\bibfnamefont {T.~W.}\ \bibnamefont
  {Clark}}, \bibinfo {author} {\bibfnamefont {R.~F.}\ \bibnamefont {Offer}},
  \bibinfo {author} {\bibfnamefont {S.}~\bibnamefont {Franke-Arnold}}, \bibinfo
  {author} {\bibfnamefont {A.~S.}\ \bibnamefont {Arnold}}, \ and\ \bibinfo
  {author} {\bibfnamefont {N.}~\bibnamefont {Radwell}},\ }\href {\doibase
  10.1364/OE.24.006249} {\bibfield  {journal} {\bibinfo  {journal} {Opt.
  Express}\ }\textbf {\bibinfo {volume} {24}},\ \bibinfo {pages} {6249}
  (\bibinfo {year} {2016})}\BibitemShut {NoStop}%
\bibitem [{\citenamefont {Vaity}\ \emph {et~al.}(2013)\citenamefont {Vaity},
  \citenamefont {Banerji},\ and\ \citenamefont {Singh}}]{VAITY20131154}%
  \BibitemOpen
  \bibfield  {author} {\bibinfo {author} {\bibfnamefont {P.}~\bibnamefont
  {Vaity}}, \bibinfo {author} {\bibfnamefont {J.}~\bibnamefont {Banerji}}, \
  and\ \bibinfo {author} {\bibfnamefont {R.}~\bibnamefont {Singh}},\ }\href
  {\doibase https://doi.org/10.1016/j.physleta.2013.02.030} {\bibfield
  {journal} {\bibinfo  {journal} {Physics Letters A}\ }\textbf {\bibinfo
  {volume} {377}},\ \bibinfo {pages} {1154 } (\bibinfo {year}
  {2013})}\BibitemShut {NoStop}%
\bibitem [{\citenamefont {Debuisschert}\ \emph {et~al.}(1993)\citenamefont
  {Debuisschert}, \citenamefont {Sizmann}, \citenamefont {Giacobino},\ and\
  \citenamefont {Fabre}}]{debuisschert_type-ii_1993}%
  \BibitemOpen
  \bibfield  {author} {\bibinfo {author} {\bibfnamefont {T.}~\bibnamefont
  {Debuisschert}}, \bibinfo {author} {\bibfnamefont {A.}~\bibnamefont
  {Sizmann}}, \bibinfo {author} {\bibfnamefont {E.}~\bibnamefont {Giacobino}},
  \ and\ \bibinfo {author} {\bibfnamefont {C.}~\bibnamefont {Fabre}},\ }\href
  {\doibase 10.1364/JOSAB.10.001668} {\bibfield  {journal} {\bibinfo  {journal}
  {Journal of the Optical Society of America B}\ }\textbf {\bibinfo {volume}
  {10}},\ \bibinfo {pages} {1668} (\bibinfo {year} {1993})}\BibitemShut
  {NoStop}%
\bibitem [{\citenamefont {Siegman}(1986)}]{siegman1986lasers}%
  \BibitemOpen
  \bibfield  {author} {\bibinfo {author} {\bibfnamefont {A.}~\bibnamefont
  {Siegman}},\ }\href {https://books.google.com.br/books?id=1BZVwUZLTkAC}
  {\emph {\bibinfo {title} {Lasers}}}\ (\bibinfo  {publisher} {University
  Science Books},\ \bibinfo {year} {1986})\BibitemShut {NoStop}%
\bibitem [{\citenamefont {Haken}(1975)}]{haken}%
  \BibitemOpen
  \bibfield  {author} {\bibinfo {author} {\bibfnamefont {H.}~\bibnamefont
  {Haken}},\ }\href {\doibase 10.1103/RevModPhys.47.67} {\bibfield  {journal}
  {\bibinfo  {journal} {Rev. Mod. Phys.}\ }\textbf {\bibinfo {volume} {47}},\
  \bibinfo {pages} {67} (\bibinfo {year} {1975})}\BibitemShut {NoStop}%
\bibitem [{\citenamefont {Lotka}(1926)}]{lotka}%
  \BibitemOpen
  \bibfield  {author} {\bibinfo {author} {\bibfnamefont {A.~J.}\ \bibnamefont
  {Lotka}},\ }\href {http://www.jstor.org/stable/43430362} {\bibfield
  {journal} {\bibinfo  {journal} {Science Progress in the Twentieth Century
  (1919-1933)}\ }\textbf {\bibinfo {volume} {21}},\ \bibinfo {pages} {341}
  (\bibinfo {year} {1926})}\BibitemShut {NoStop}%
\bibitem [{\citenamefont {Volterra}(1931)}]{volterra}%
  \BibitemOpen
  \bibfield  {author} {\bibinfo {author} {\bibfnamefont {V.}~\bibnamefont
  {Volterra}},\ }\href
  {http://www.gabay-editeur.com/VOLTERRA-Lecons-sur-la-theorie-mathematique-de-la-lutte-pour-la-vie-1931}
  {\emph {\bibinfo {title} {Leçons sur la théorie mathématique de la lutte
  pour la vie}}}\ (\bibinfo  {publisher} {J. Gabay (Paris)},\ \bibinfo {year}
  {1931})\BibitemShut {NoStop}%
\bibitem [{\citenamefont {Goel}\ \emph {et~al.}(1971)\citenamefont {Goel},
  \citenamefont {Maitra},\ and\ \citenamefont {Montroll}}]{Goel:1971}%
  \BibitemOpen
  \bibfield  {author} {\bibinfo {author} {\bibfnamefont {N.~S.}\ \bibnamefont
  {Goel}}, \bibinfo {author} {\bibfnamefont {S.~C.}\ \bibnamefont {Maitra}}, \
  and\ \bibinfo {author} {\bibfnamefont {E.~W.}\ \bibnamefont {Montroll}},\
  }\href {\doibase 10.1103/RevModPhys.43.231} {\bibfield  {journal} {\bibinfo
  {journal} {Rev. Mod. Phys.}\ }\textbf {\bibinfo {volume} {43}},\ \bibinfo
  {pages} {231} (\bibinfo {year} {1971})}\BibitemShut {NoStop}%
\bibitem [{\citenamefont {Sargent}\ \emph {et~al.}(1974)\citenamefont
  {Sargent}, \citenamefont {Lamb},\ and\ \citenamefont
  {Scully}}]{sargent1974laser}%
  \BibitemOpen
  \bibfield  {author} {\bibinfo {author} {\bibfnamefont {M.}~\bibnamefont
  {Sargent}}, \bibinfo {author} {\bibfnamefont {W.}~\bibnamefont {Lamb}}, \
  and\ \bibinfo {author} {\bibfnamefont {M.}~\bibnamefont {Scully}},\ }\href
  {https://books.google.com.br/books?id=4f3DtAEACAAJ} {\emph {\bibinfo {title}
  {Laser Physics}}}\ (\bibinfo  {publisher} {Addison-Wesley Publishing Company,
  Advanced Book Program},\ \bibinfo {year} {1974})\BibitemShut {NoStop}%
\end{thebibliography}%

\end{document}